\documentclass[12pt]{spieman}  
\usepackage{amsmath,amsfonts,amssymb}
\usepackage{graphicx}
\usepackage{setspace}
\usepackage{tocloft}
\usepackage{lineno}
\usepackage[version=4]{mhchem}
\usepackage{float}
\usepackage{svg}
\usepackage{upgreek}
\usepackage{makecell}

\title{Science-Driven Performance Requirements for UV Coronagraph Design: Implications for \ce{O3} Detection in Earth-like Atmospheres}

\author[a,b]{Amber V. Young}
\author[b]{Giada Arney}
\author[a,b]{Shawn Domagal-Goldman}
\author[c,d,e]{Tyler D. Robinson}
\author[b]{Roser Juanola-Parramon}
\author[b]{Christopher Stark}
\author[f,g,h,i]{Renyu Hu}
\author[i]{Armen Tokadjian}
\author[i]{Mario Damiano}
\affil[a]{Mary W. Jackson NASA Headquarters, Washington, DC, USA}
\affil[b]{NASA Goddard Space Flight Center, Greenbelt, MD, 20771}
\affil[c]{Habitability, Atmospheres, and Biosignatures Laboratory, University of Arizona, Tucson, AZ 85721, USA}
\affil[d]{NASA Nexus for Exoplanet System Science Virtual Planetary Laboratory, SETI Institute, Mountain View, CA 94043, USA}
\affil[e]{Lunar \& Planetary Laboratory, University of Arizona, Tucson, AZ 85721 USA}
\affil[f]{Department of Astronomy \& Astrophysics, The Pennsylvania State University, University Park,
PA 16802, USA}
\affil[g]{Center for Exoplanets and Habitable Worlds, The Pennsylvania State University, University Park,
PA 16802, USA}
\affil[h]{Institute for Computational and Data Science, The Pennsylvania State University, University Park,
PA 16802, USA}
\affil[i]{Jet Propulsion Laboratory, California Institute of Technology, Pasadena, CA 91109, USA}

\cftpagenumbersoff{figure}
\cftpagenumbersoff{table} 
\begin{document} 
\maketitle

\begin{abstract}
 Detecting biosignatures on Earth-like exoplanets from the Proterozoic era requires instrumentation capable of probing near-ultraviolet (NUV) spectral features, particularly ozone (\ce{O3}), which serves as an indirect proxy for molecular oxygen (\ce{O2}) when direct detection is challenging. Using a combination of photochemical-climate modeling and atmospheric retrieval simulations, we quantify fiducial instrument performance requirements for robust \ce{O3} detection and assess the risk of spectral false positives from sulfur dioxide (\ce{SO2}). Our analysis demonstrates the \ce{O3} abundances for Proterozoic Earth-like planets with \ce{O2} as low as 0.1\% the present atmospheric level, can be constrained to within an order of magnitude at 1\,$\sigma$ confidence using three adjacent photometric bandpasses at a signal-to-noise ratio of 10 at 0.366\,$\upmu$m, each with a 15\% width, and a short wavelength cutoff for those observations near 0.26\,$\upmu$m. The self-consistent steady state atmospheric modeling results show that \ce{SO2} buildup is difficult in a habitable environment with appreciable amounts of water vapor in its atmosphere and would require volcanic outgassing rates at least two orders of magnitude above present-day Earth to reach surface abundance levels comparable to Proterozoic Earth-like abundances of \ce{O3}. Even under these high \ce{SO2} flux scenarios, \ce{SO2} remains largely unconstrained with the recommended observational configuration, reinforcing the need for NUV coverage in coronagraphic instruments. Our findings support the inclusion of UV-optimized optics and detectors on the Habitable Worlds Observatory (HWO) to enable biosignature detection strategies sensitive to atmospheric conditions across key phases of Earth's geological history.
\end{abstract}


{\noindent \footnotesize\textbf{*}Amber V. Young,  \linkable{amber.v.young@nasa.gov} }

\begin{spacing}{2}   

\section{Introduction}
\label{sec:intro}

Detecting ozone (\ce{O3}) in the near-ultraviolet (NUV) is central to characterizing potentially Earth-like planets because the Hartley-Huggins bands ($\sim$0.200\,--\,0.350\,$\upmu$m) are strong spectral absorption features that remain a diagnostic even for weakly oxygenated conditions, as was the case for a substantial fraction of Earth's history. Thus, observational coverage of the NUV is key for future biosignature searches with the Habitable Worlds Observatory (HWO). Despite its scientific merit, high-contrast coronagraphy in the UV \cite{Gorkom_etal_2025, Juanola_Parramon_2026} has not yet been demonstrated at the required performance levels. Advancements to maturing starlight suppression are on the horizon with technology demonstrations from the Roman Space Telescope, which is set to launch in 2026 but will primarily focus on the visible wavelength range \cite{Bailey_RST_2023}. Prior mission concept studies like the Large UV/Optical/Infrared\cite{LUVOIR_Paper_2018} (LUVOIR) ECLIPS coronagraph \cite{pueyo2019luvoir, juanola2019luvoir} and the Habitable Exoplanet Observatory starshade \cite{HabEx_Paper_2018} have explored technology considerations in the UV but continuing to assess and conduct performance trades is needed.

Advancements in this field are motivated by results from the previous Decadal Survey on Astronomy and Astrophysics, which recommended prioritizing a mission capable of conducting a search for life in the UV/optical/near-infrared wavelength range, and such a mission later arose as HWO \cite{NAS_2021}. One of HWO's primary mission objectives is to search for life on a target sample of $\sim$25 potentially habitable candidates. In order to guide observatory requirements regarding the search for life, Earth is a natural starting point for designing an observatory capable of characterizing the only known example of a life bearing planet. Over the course of Earth's geological history, Earth has drastically evolved, and its atmospheric state has been strongly influenced by the presence of biology. One such geological time period was the Proterozoic era: it represents roughly half of Earth's geological history (from $\sim$2.5 to 0.54 billion years ago) and encompasses some of Earth's most drastic changes including the atmospheric transition from an anoxic to oxygenated state following the Great Oxidation Event. \ce{O2} levels were relatively low during this period of time compared to the modern day, with model predictions suggesting abundances as low as 0.1\% of the present atmospheric level of \ce{O2} \cite{Planavsky_2014}.

From a characterization perspective, \ce{O2} has long been a key atmospheric species associated with life detection because it is a main byproduct of biologically driven processes such as photosynthesis. While an oxygen rich biosphere such as a modern Earth could be contextualized with observations at visible wavelengths via strong absorption from the \ce{O2}-A band at 0.76\,$\upmu$m and the \ce{O3} Chappuis band that extends from 0.5\,$\upmu$m to 0.7\,$\upmu$m, constraining the reduced abundance of \ce{O2} associated with the Proterozoic era would be much more challenging in the UV to near infrared \cite{Schwieterman_2018}.  However, \ce{O3} can remain detectable even at reduced \ce{O2} abundances in part due to the sheer strength of \ce{O3} absorption in the NUV, making these observations essential for characterizing the presence of oxygen-bearing species in a Proterozoic-like planet’s atmosphere \cite{Reinhard_2017,Damiano_2023}. Given that there are numerous barriers to oxygen accumulation in a planetary atmosphere \cite{Lyons_2014}, low oxygen exoplanets akin to the Proterozoic Earth may be more common than richly-oxygenated planets akin to modern Earth, making access to the NUV \ce{O3} feature critical for characterizing such worlds. \ce{O3} itself is also a key atmospheric species to constrain on other worlds given its ability to provide UV shielding and maintaining hospitable conditions on the surface of a planet. Alongside constraining the atmospheric abundance of \ce{O3} it is also critical to take into account potential spectral false positives which include chemical species that have overlapping spectral absorption features with \ce{O3} that could mimic \ce{O3} absorption. \ce{SO2} is a prime example of a chemical species that has an overlapping absorption feature with \ce{O3} in the NUV and determining our ability to distinguish between the two species is another key consideration incorporated in this analysis. 

Atmospheric characterization is typically performed using statistical inference methods that fit forward models to observational data, most often through radiative transfer modeling of planetary spectra. These retrieval techniques have been widely applied to extract key atmospheric properties such as surface temperature, pressure, and opacity sources including molecular absorption, clouds, and hazes from observed spectra \cite{Madhusudhan_2009,Benneke_2012,Line_2013,Lupu_2016,Feng_2018,Barstow_2020,MacDonald_2023}. Characterizing the oxygenation of planetary targets is a strong science motivator for development of UV coronagraphic technology and in this work, we used atmospheric inference through Bayesian statistical framework tools in order to translate \ce{O3} detectability into science-driven performance requirements for a UV coronagraph in terms of short wavelength cutoff, bandpass placement, and observational signal-to-noise ratio (SNR), needed to achieve robust constraints on the \ce{O3} abundance for Proterozoic Earth-like atmospheres. These results provide actionable inputs for UV coronagraph design and testbed planning, bridging science needs and engineering feasibility for future Habitable Worlds Observatory mission planning and development. This work expands upon preliminary findings reported in ``Towards Habitable Worlds Observatory: Visionary Science and Transformational Technology" proceedings (in press), providing expanded atmospheric modeling, retrieval simulations and associated discussion of new results \cite{Young_2026}. 

\section{The Atmospheric Model} 
\label{sec:Atmos}

\texttt{Atmos} is a one dimensional photochemical-climate coupled model that simulates the chemical composition and thermal structure of planetary atmospheres \cite{arney_pale_2016,Arney_pale_2017}. It self consistently computes the atmospheric state given planetary parameters (e.g., gravity, radius, stellar insolation), stellar parameters (wavelength dependent flux), and atmospheric inputs (chemical species profiles, chemical species fluxes). The model iteratively solves for a steady state solution, producing altitude-dependent profiles of species mixing ratios, haze properties, temperature and pressure. By tracking the production and loss rates of each species, \texttt{Atmos} adaptively increases the time-steps, enabling convergence toward a stable solution.

\section{The Retrieval Model} 
\label{sec:rfast}

\texttt{rfast} is an atmospheric retrieval tool that combines three core components: a radiative transfer forward model, an instrument noise simulator, and a Bayesian framework implemented using the Markov Chain Monte Carlo (MCMC) technique \cite{Robinson_2023}. The forward model simulates planetary spectra in 1D or 3D geometries for reflected light, transmission, and emission, incorporating Rayleigh scattering, molecular opacity absorption, and thermal emission via a computationally efficient two-stream radiative transfer approximation. Its accuracy has been validated against other radiative transfer models like SMART \cite{Meadows_1996} and tested against diverse datasets including Earth reflectance spectra from the NASA EPOXI mission \cite{Livengood_2011}, and Titan transit spectra \cite{Robinson_2014} derived from Cassini observations \cite{Bellucci_2009,Hayne_2014,Maltagliati_2015}. The instrument noise simulator generates synthetic spectra based on user-defined telescope parameters (e.g., wavelength range, resolving power, specified SNR) and applies user-selected noise treatments (e.g., constant error, constant SNR). Retrievals are performed using the open source Python package \texttt{emcee} \cite{Foreman_Mackey_2013}, which computes posterior probabilities that are calculated using the likelihood, user specified free parameters, set ranges for each parameter's prior, and the type of prior (Gaussian or uniform).

\section{Atmospheric Modeling/Simulation Setup} 
\label{sec:modeling}

The \texttt{Atmos} boundary conditions were configured to simulate a range of \ce{SO2} outgassing scenarios, spanning from low modern day fluxes ($9 \times 10^{9}$\,molec\,cm$^{-2}$\,sec$^{-1}$) to high Archean Earth fluxes approximately 100 times greater than present-day values ($9 \times 10^{11}$\,molec\,cm$^{-2}$\,sec$^{-1}$), following estimates from \cite{Zahnle_2006}. The boundary conditions for \ce{SO2} were set to a constant upward flux at the surface paired with a constant deposition velocity set to the model default of 1\,$cm s^{-1}$, representing dry deposition at the surface \cite{Zahnle_2006}. To mimic the vertical distribution of volcanic emissions, \ce{SO2} was allowed to mix into the troposphere of the atmosphere. This vertical extent was applied uniformly across all flux scenarios. The model was run in photochemical-climate coupled mode, iteratively solving for steady-state photochemistry and using the resulting species abundances to inform radiative feedbacks and compute a temperature-pressure (T-P) profile. This updated T-P profile structure was then fed back into the photochemical model, allowing for dynamic coupling between atmospheric composition and climate. To reach a stable solution, we adopted a short time-stepping approach where the photochemical (Photo) and climate (Clima) models are run sequentially in an iterative loop \cite{Teal_2022}. Clima is intentionally limited to a small number of iterations per coupling cycle to prevent full adjustment of radiative forcing from spectrally active chemical species, which can otherwise lead to oscillations between non-self-consistent states. Over hundreds of coupling iterations, this approach enables convergence to a more reliable steady-state solution in radiative-convective equilibrium. 

\begin{figure} [H]
\begin{center}
\begin{tabular}{c}
\includegraphics[height=8cm]{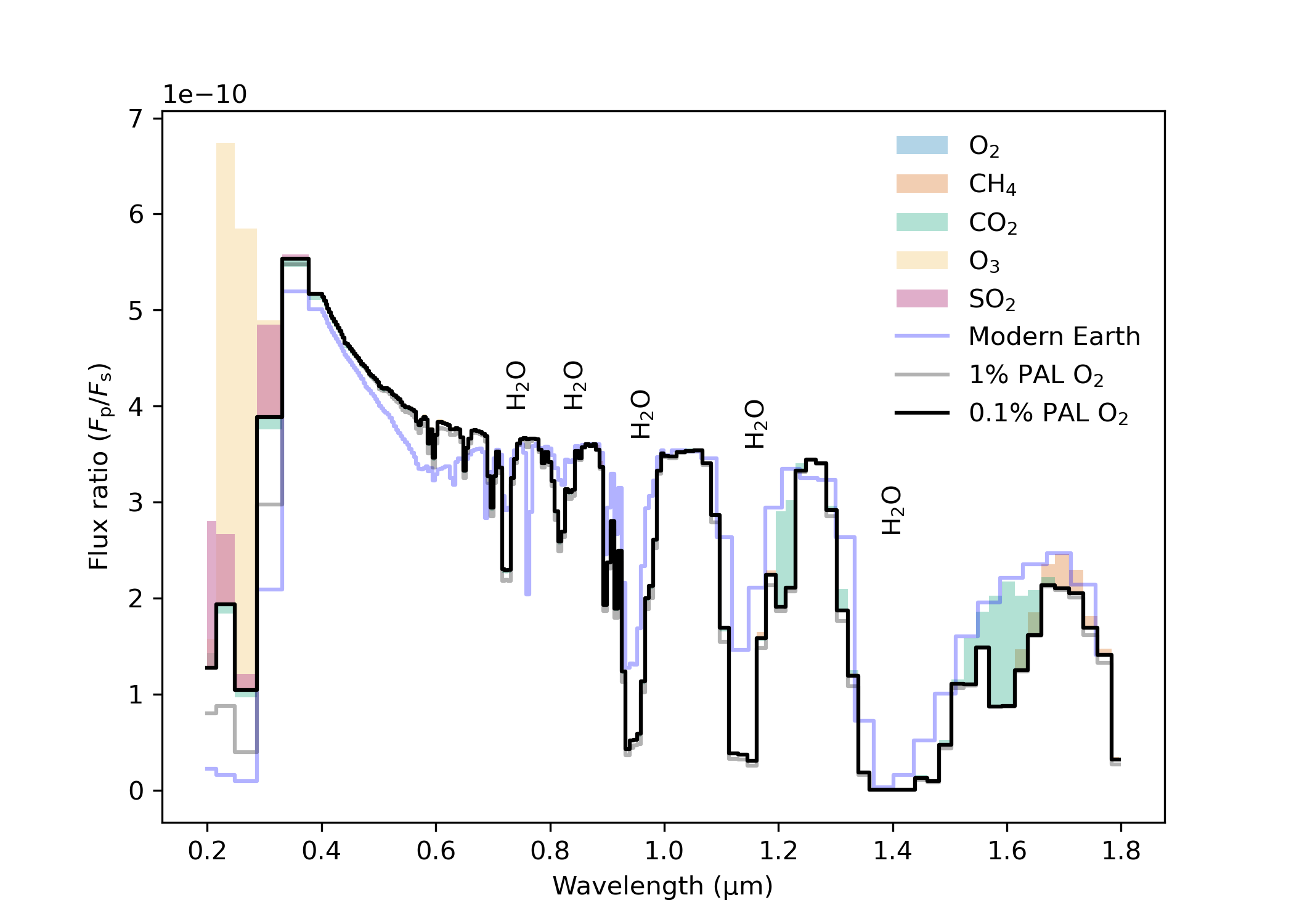}
\end{tabular}
\end{center}
\caption 
{ \label{Spectral_Contrib}
Simulated reflected light spectra of both Proterozoic Earth cases (0.1\% and 1\% present atmospheric level (PAL) of \ce{O2}) with the modern Earth spectrum also plotted for reference. The spectral contribution of each chemical species is indicated with color coded shaded regions.} 
\end{figure}

Simulations were initialized with Proterozoic Earth-like conditions, focusing on the low end of predicted \ce{O2} abundances (0.1 and 1\% the present atmospheric level), elevated \ce{CH4} levels (100 - 300$\times$ the modern abundance), and \ce{N2} as the dominant background gas, consistent with previous model predictions \cite{Pavlov_2003, Planavsky_2014}. This regime was selected to assess the detectability of \ce{O3} under more challenging conditions where its abundance is expected to be minimal. Planetary parameters such as surface gravity, radius, orbital distance, and total surface pressure were fixed at Earth-like values. The input stellar spectrum was scaled to reflect the reduced solar luminosity 1 billion years ago (at 1.0 Ga) \cite{Claire_2012}. Figure \ref{Spectral_Contrib} depicts the reflected light spectra for each of the atmospheric cases and the spectral absorption of each chemical species included as a free parameter in the retrievals. 

We used \texttt{rfast} to generate forward model spectra spanning 0.2 - 0.45\,$\upmu$m, using the temperature, pressure, and atmospheric abundance profiles calculated with \texttt{Atmos}. \ce{N2} was used as a filler gas, which ensured that the total mixing ratio of chemical species abundances always summed to unity. The forward model was run with 50 atmospheric layers consistent with default model settings. Species opacities were derived from the HITRAN database \cite{Gordon_2022_HITRAN} using the Line-By-Line ABSorption Coefficients tool \cite{Meadows_1996} and generally this setup follows the forward modeling approach outlined in \cite{Robinson_2023}. The signal-to-noise ratio (SNR) was defined relative to the spectral brightness at 0.366\,$\upmu$m, with error bars scaled accordingly and applied uniformly across the spectral data, independent of the source flux at a level of $\sim 5 \times 10^{-11}$. This is representative of a noise regime that is dominated by systematic noise. The uncertainties on the simulated datasets produced here are in line with a recent study, which estimated a required noise floor of at least $3 \times 10^{-11}$ to enable detections of Earth-like planets \cite{Nemati_2020}.

Each retrieval used the \texttt{emcee} MCMC sampler with 200 walkers run for 200,000 steps. The first 150,000 steps were discarded as burn-in to allow the chains to reach stable, converged solutions. Convergence was evaluated using trace plots of walker positions across iterations, confirming that chains were sufficiently converged at the chosen burn-in and retained samples accurately reflected the converged parameter space. The simulated noisy spectrum was held fixed on the truth spectrum in all retrieval cases. In some previous work, synthetic datasets were generated using multiple independent noise instances, with separate MCMC chains run for each and the resulting samples combined to form the final posterior \cite{Feng_2018,Young_2024,Young_Decision_Tree_2024,Young_2025}. This practice can introduce parameter biases, arising from residual noise in the data spectra \cite{Feng_2018,Konrad_2022}. For this work, we therefore adopt spectra with fixed noise centered on the truth spectrum, consistent with feasibility studies, in order to focus on required SNR and parameter sensitivity and manage overall computational cost.

To assess the impact of spectral coverage on \ce{O3}
detectability, we performed a sequence of retrievals in which the short-wavelength cutoff was incrementally increased. The full spectral range was divided into evenly spaced bandpasses with a resolving power ($\lambda/\Delta \lambda$) of 5 representing low-resolution spectroscopy. This allowed us to identify the minimum spectral range required to robustly constrain \ce{O3} in the presence of potential false positive signals from \ce{SO2}.

In a separate set of simulations, we used the photometry mode in \texttt{rfast} to evaluate the performance of specific bandpass configurations in distinguishing \ce{O3} from \ce{SO2}. Four configurations were tested: \textbf{1A}: two adjacent 22\% bandpasses, \textbf{2B}: three adjacent 15\% bandpasses, \textbf{3A}: three non-contiguous 10\% bandpasses, and \textbf{3B}: four adjacent 10\% bandpasses. These configuration choices were motivated by prior studies, including laboratory experiments that informed bandpass recommendations \cite{Mennesson_2024}, and planetary yield calculations that assessed observational performance across mission scenarios \cite{Morgan_2024}. Each configuration was evaluated at SNRs of 3, 5, and 10 to explore detectability across a range of observational data qualities.

All of the retrieved cases included a suite of 14 retrieved parameters (listed in Table \ref{tab_retrieval_params} including chemical abundances (\ce{O2}, \ce{H2O}, \ce{CO2}, \ce{O3}, \ce{CH4}, \ce{SO2}), thermodynamic parameters (surface pressure), planetary parameters (surface albedo, radius, gravity), and cloud parameters (vertical extent, the cloud top pressure, optical depth, and cloud fraction). Assumed prior ranges for each parameter are provided in Table \ref{tab_retrieval_params} as well. Notably for the chemical species, we adopt broad uniform priors designed to allow for exploration from trace to bulk abundances within strict physical limits. The exceptions are \ce{O3} and \ce{SO2} whose upper bounds are reduced to reflect their role as photochemically produced trace species that cannot physically reach bulk-gas abundances in the atmospheric regimes considered.


\begin{table}[H]
    \centering
    \begin{tabular}{l c c}
    \hline
    \hline
         Parameter & Description & Prior  \\
         \hline
         log $P_0$ (log Pa) & Surface pressure & [0,8]\\
         log O$_2$ & Molecular oxygen mixing ratio & [-10,0]\\
         log H$_2$O & Water vapor mixing ratio & [-10,0]\\
         log CO$_2$ & Carbon dioxide mixing ratio& [-10,0]\\
         log O$_3$ & Ozone mixing ratio&[-15,-2]\\
         log CH$_4$ & Methane mixing ratio&[-10,0]\\
         log SO$_2$ & Methane mixing ratio&[-20,-2]\\
         log $A\rm_s$ & Surface albedo&[-2,0]\\
         log $R\rm_p$ (log R$_\oplus$) & Planetary Radius &[-1,1]\\
         log g (log $\rm{m/s^{2}}$) & Planetary gravity&[0,1.39]\\
         log $\Delta p \rm_c$ (log Pa) & Cloud thickness &[0,8]\\
         log $p\rm_t$ (log Pa) & Cloud top pressure&[0,8]\\
         log $\tau\rm_c$ & Cloud optical depth&[-3,3]\\
         log $f\rm_c$ & Cloud fraction&[-3,0]\\
         \hline
         \hline
    \end{tabular}
    \caption{List of planetary parameters that were retrieved in this work using  \texttt{rfast}. Listed are the parameter names, descriptions, and their corresponding priors (which are flat across the indicated range).}
    \label{tab_retrieval_params}
\end{table}

\begin{figure}
\begin{center}
\begin{tabular}{c}
\includegraphics[height=8cm]{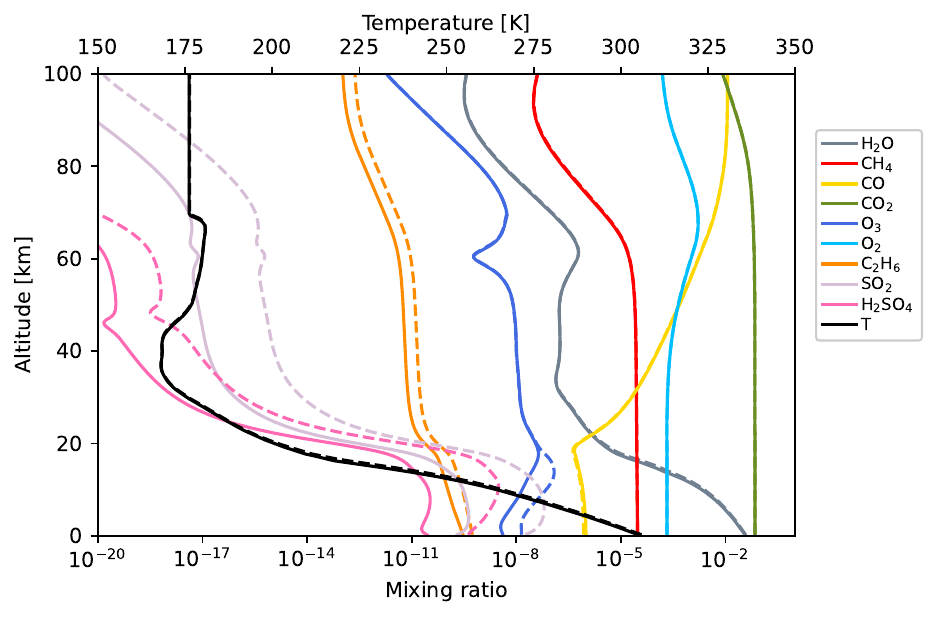}
\end{tabular}
\end{center}
\caption 
{ \label{PE0_Profiles_S_species}
tmospheric profiles of key chemical species (color coded lines) and temperature (black line) from self consistent steady-state modeling with \texttt{Atmos} for 0.1\% the present atmospheric level of \ce{O2}. Solid lines show profiles using the default \ce{SO2} flux boundary condition; dashed lines show profiles using the high \ce{SO2} flux boundary condition.} 
\end{figure}

\begin{figure} [H]
\begin{center}
\begin{tabular}{c}
\includegraphics[height=8cm]{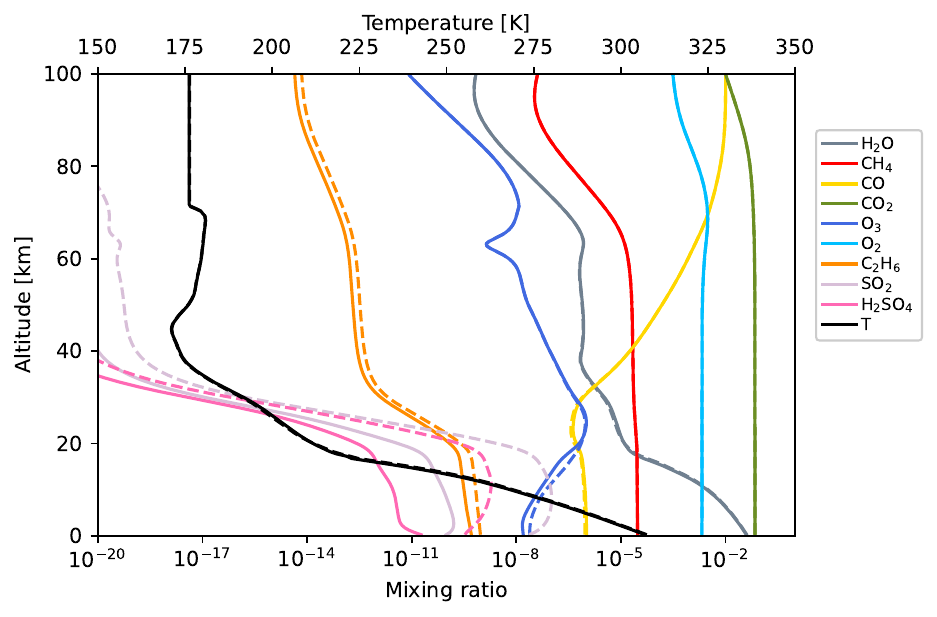}
\end{tabular}
\end{center}
\caption 
{ \label{PE_Profiles_S_species}
Atmospheric profiles of key chemical species (color coded lines) and temperature (black line) from self consistent steady-state modeling with \texttt{Atmos} for 1\% the present atmospheric level of \ce{O2}. Solid lines show profiles using the default \ce{SO2} flux boundary condition; dashed lines show profiles using the high \ce{SO2} flux boundary condition.} 
\end{figure}

The altitude dependent profiles for chemical species \ce{H2O}, \ce{CH4}, \ce{CO}, \ce{CO2}, \ce{O3}, \ce{O2}, \ce{C2H6}, \ce{SO2}, \ce{H2SO4} alongside the temperature structure are presented in Figure \ref{PE0_Profiles_S_species} for the 0.1\% present level of \ce{O2} atmospheric scenario and Figure \ref{PE_Profiles_S_species} for the 1\% present level of \ce{O2} atmospheric scenario. Solid lines correspond to self-consistent steady-state profiles assuming a surface \ce{SO2} outgassing flux of $9 \times 10^{9}$\,molec\,cm$^{-2}$\,sec$^{-1}$, while the dashed lines reflect a higher flux scenario of $9 \times 10^{11}$\,molec\,cm$^{-2}$\,sec$^{-1}$. At elevated fluxes, surface \ce{SO2} abundance approaches that of the \ce{O3}, though the vertical profiles of most other species remain largely unchanged. Notably the \ce{H2SO4} exhibits a substantial increase across all altitudes in response to the enhanced sulfur content, consistent with its photochemical production pathway.

The column average mixing ratios for all species included in the retrieval are reported in Table \ref{atm_species} for each of the atmospheric scenarios.

\begin{table}[H]
    \centering
    \begin{tabular}{l c c c c}
    \hline
    \hline
        Species & 
        \makecell{1\% PAL \ce{O2} \\ High \ce{SO2}} & 
        \makecell{1\% PAL \ce{O2}\\ Default \ce{SO2}} & 
        \makecell{0.1\% PAL \ce{O2}\\ High \ce{SO2}} & 
        \makecell{0.1\% PAL \ce{O2}\\ Default \ce{SO2}}\\
    \hline
    \hline
        \ce{H2O} & $1.37\times 10^{-2}$ & $1.33 \times 10^{-2}$ & $1.23 \times 10^{-2}$ & $1.16 \times 10^{-3}$\\
        \ce{CH4} & $2.95 \times 10^{-5}$ & $2.94 \times 10^{-5}$ & $2.95 \times 10^{-5}$ & $2.93 \times 10^{-5}$\\
        \ce{CO2} & $7.00 \times 10^{-2}$ & $7.00 \times 10^{-2}$ & $7.00 \times 10^{-2}$ & $7.00 \times 10^{-2}$\\
        \ce{O2} & $2.10 \times 10^{-3}$ & $2.10 \times 10^{-3}$ & $2.09 \times 10^{-4}$ & $2.1 \times 10^{-4}$\\
        \ce{O3} & $1.37 \times 10^{-7}$ & $1.06 \times 10^{-7}$ & $4.05 \times 10^{-8}$ & $1.24 \times 10^{-8}$\\
        \ce{SO2} & $6.67 \times 10^{-8}$ & $5.62 \times 10^{-10}$ & $3.90 \times 10^{-8}$ & $2.90 \times 10^{-10}$\\
    \hline
    \hline
\end{tabular}
    \caption{Column-averaged mixing ratios for key atmospheric species across the four atmospheric scenarios. Each column corresponds to a distinct atmospheric boundary condition case: \ce{O2} fixed at either 1\% or 0.1\% the present atmospheric level (PAL), combined with either a high \ce{SO2} outgassing flux or the default \ce{SO2} flux used in \texttt{Atmos}. All values are dimensionless mixing ratios.}
    \label{atm_species}
\end{table}

\section{Results} 
\label{sec:results}
\begin{figure}
\begin{center}
\begin{tabular}{c}
\includegraphics[height=6.5cm]{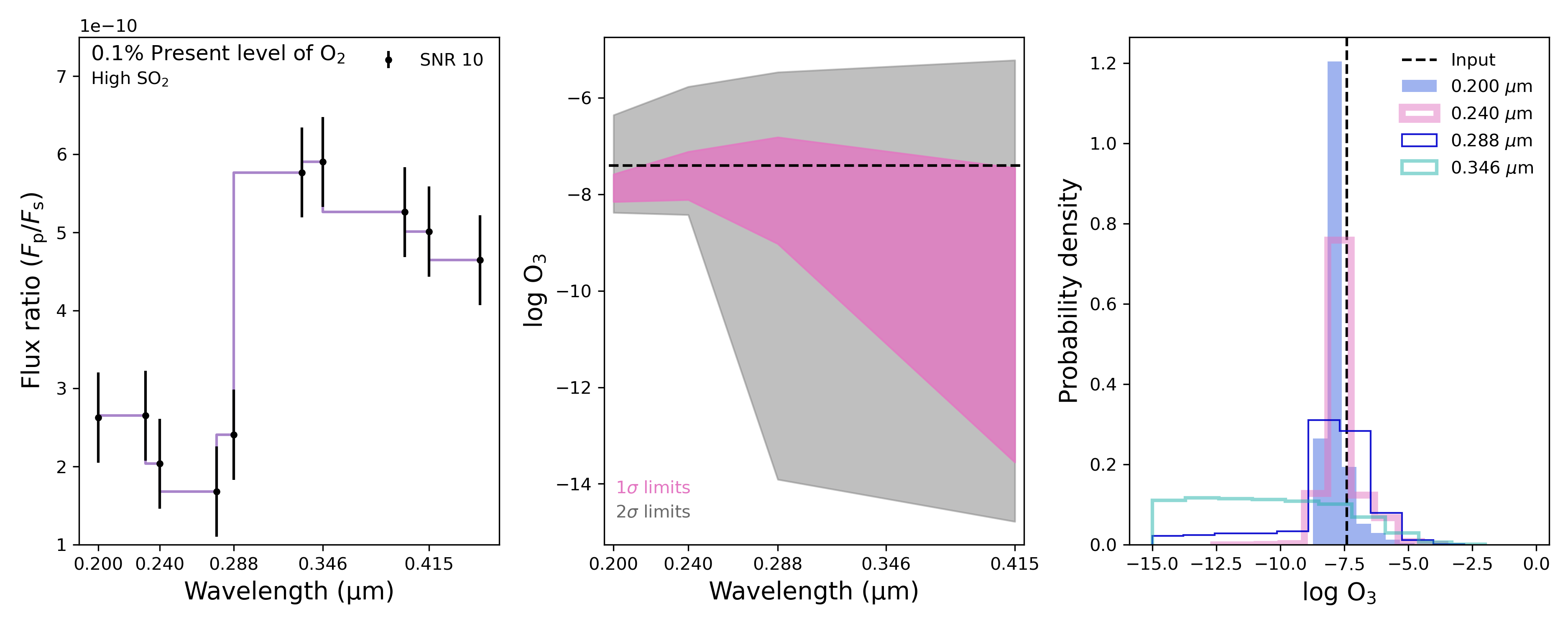}
\end{tabular}
\end{center}
\caption 
{ \label{swl_PE0_HighSO2}
Retrieval results for \ce{O3} abundance as a function of short wavelength cutoff under a high \ce{SO2} flux scenario with 0.1\% present day \ce{O2}. Left: Forward-model spectrum with simulated data (SNR=10 set relative to the flux at 0.366\,$\upmu$m); the full range is divided into 20\% bandpasses (R$\approx$7). Middle: 1\,$\sigma$ (pink) and 2\,$\sigma$ (grey) credible regions versus short wavelength cutoff. Right: Marginal posterior distributions of \ce{O3} abundance for each cutoff. The black dashed lines indicate the truth (input) column-averaged mixing ratio.} 
\end{figure}

\begin{figure}
\begin{center}
\begin{tabular}{c}
\includegraphics[height=6.5cm]{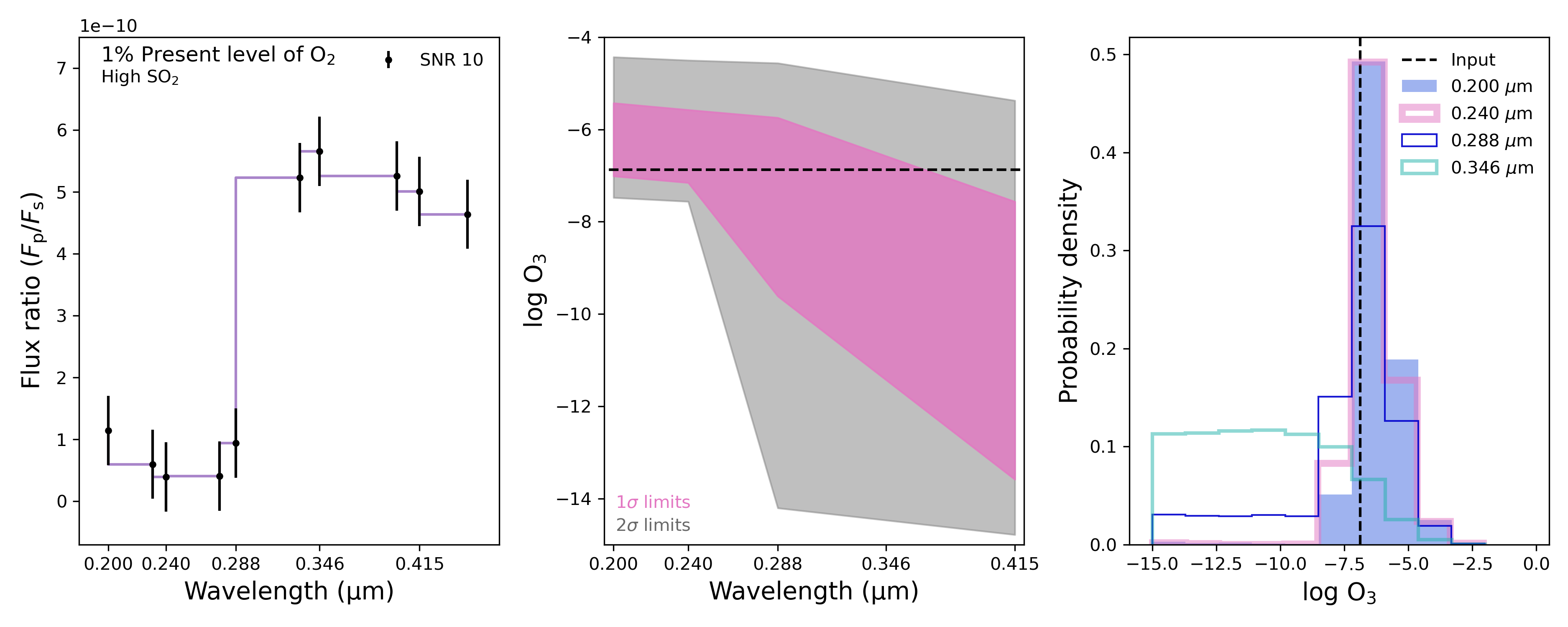}
\end{tabular}
\end{center}
\caption 
{ \label{swl_PE_HighSO2}
Retrieval results for \ce{O3} abundance as a function of short wavelength cutoff under a high \ce{SO2} flux scenario with 1\% present day \ce{O2}. Left: Forward-model spectrum with simulated data (SNR=10 set relative to the flux at 0.366\,$\upmu$m); the full range is divided into 20\% bandpasses (R$\approx$7). Middle: 1\,$\sigma$ (pink) and 2\,$\sigma$ (grey) credible regions versus short wavelength cutoff. Right: Marginal posterior distributions of \ce{O3} abundance for each cutoff. The black dashed lines indicate the truth (input) column-averaged mixing ratio.} 
\end{figure}

Figure \ref{swl_PE0_HighSO2} shows the results for the short wavelength cutoff retrievals where we sequentially increased the short wavelength cutoff to determine the minimum threshold for maintaining detections of \ce{O3}. The leftmost panel shows the simulated data at SNR 10 over plotted against the forward model spectrum. The middle panel shows the 1\,$\sigma$ and 2\,$\sigma$ credible regions for each short wavelength cutoff. We can see that at 0.288\,$\upmu$m the 2\,$\sigma$ region exhibits a lower limit constraint that is influenced by the assumed priors on the \ce{O3} abundance. At 0.346\,$\upmu$m, we are unable to constrain the \ce{O3} abundance at all due to limited coverage of the spectral slope. While the best constraints are obtained for the 0.200\,$\upmu$m cutoff, the 1\,$\sigma$ constraints are well within an order of magnitude spread out to approximately 0.26\,$\upmu$m constituting tight constraints. Similar trends can be seen for the 1\% present level \ce{O2} scenario in Figure \ref{swl_PE_HighSO2}. Constraints on \ce{O3} are more challenging to obtain in this case due to the limited flux at the deepest part of the absorption feature. However, constraints slightly exceeding an order of magnitude can be achieved when the short wavelength cutoff extends below 0.288\,$\upmu$m. Notably, the posterior distribution for the 0.288\,$\upmu$m cutoff exhibits a statistically significant tail extending toward trace abundances at an SNR of 10. 



\begin{figure}
\begin{center}
\begin{tabular}{c}
\includegraphics[height=7.5cm]{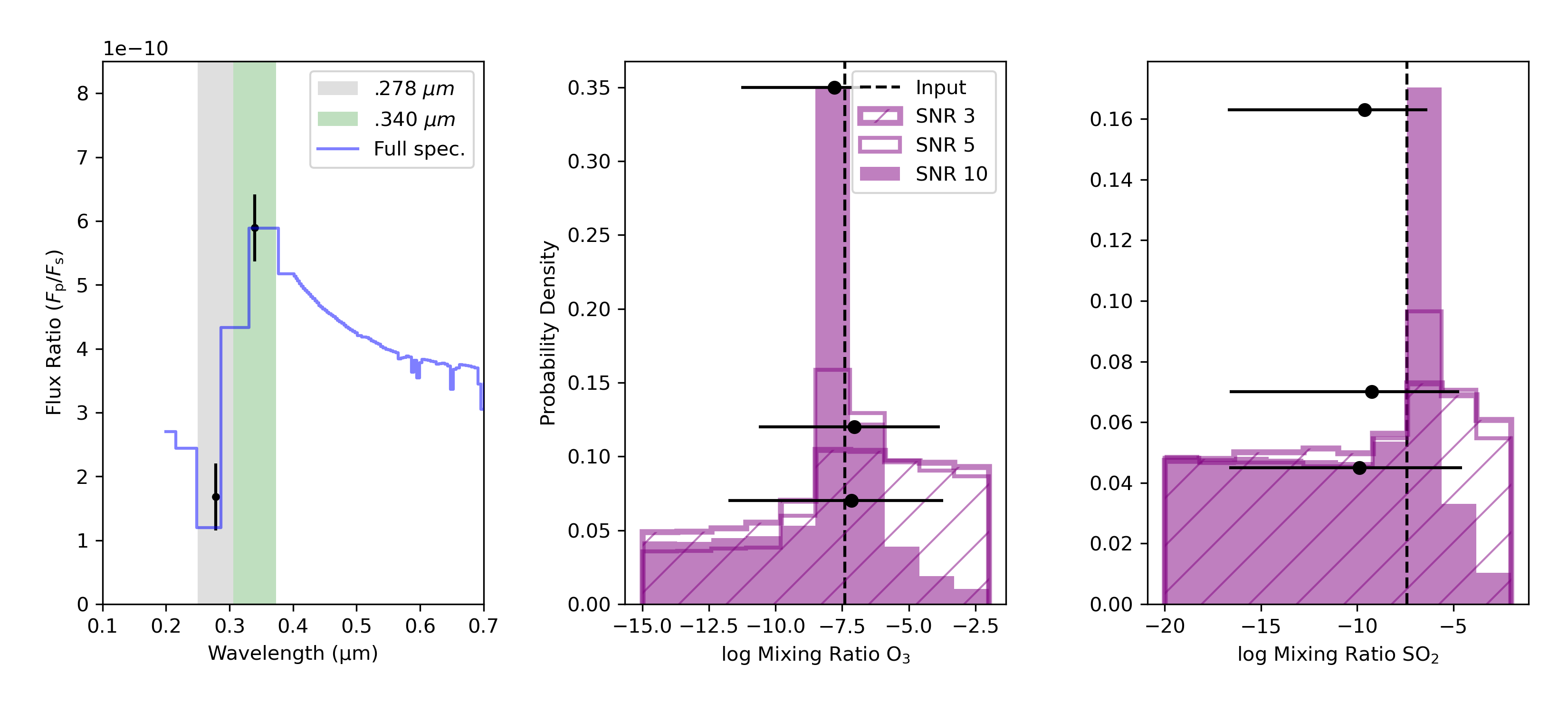}
\end{tabular}
\end{center}
\caption 
{ \label{DISRA_1A}
Retrieval results for the \textbf{1A} bandpass configuration under a high \ce{SO2} flux scenario with 0.1\% present day \ce{O2}. Left: Forward-model spectrum with simulated data with SNR=10 set relative to the flux at 0.33\,$\upmu$m shown for reference. Grey shading marks the 0.25\,--\,0.306\,$\upmu$m bandpass and green shading marks the 0.306\,--\,0.373\,$\upmu$m bandpass. Middle: Marginal posterior distributions for the \ce{O3} abundance at SNR=3 (hatched), 5 (unfilled), and 10 (filled). Right: Same as the middle panel but for the \ce{SO2} abundance. Horizontal bars above each posterior indicate the $\pm$1\,$\sigma$ regions, and the black point marks the mean value. Black vertical dashed lines indicate the truth (input) column-averaged mixing ratio abundance for each species.} 
\end{figure}

Figure \ref{DISRA_1A} shows two adjacent 22\% bandpasses spanning  0.250\,--\,0.373\,$\upmu$m. In the left panel, the forward model spectrum (blue) is overplotted with simulated photometric data points, one generated within each bandpass and centered between the bandpass limits. Shown for reference are the simulated data at SNR 10, calculated relative to the flux at 0.33\,$\upmu$m. Black horizontal lines indicate the 1\,$\sigma$ confidence widths for each distribution, with the black point marking the mean value. The center panel shows marginal posterior distributions for \ce{O3} at SNR 3 (hatched), 5 (unfilled), and 10 (filled) for the 0.1\% present level of \ce{O2} atmospheric scenario. The right panel displays marginal distributions for \ce{SO2} assuming the high flux outgassing scenario. In this case, tight constraints cannot be achieved for either species, with 1\,$\sigma$ credible regions spanning several orders of magnitude across all of the SNR cases.


\begin{figure}
\begin{center}
\begin{tabular}{c}
\includegraphics[height=7.5cm]{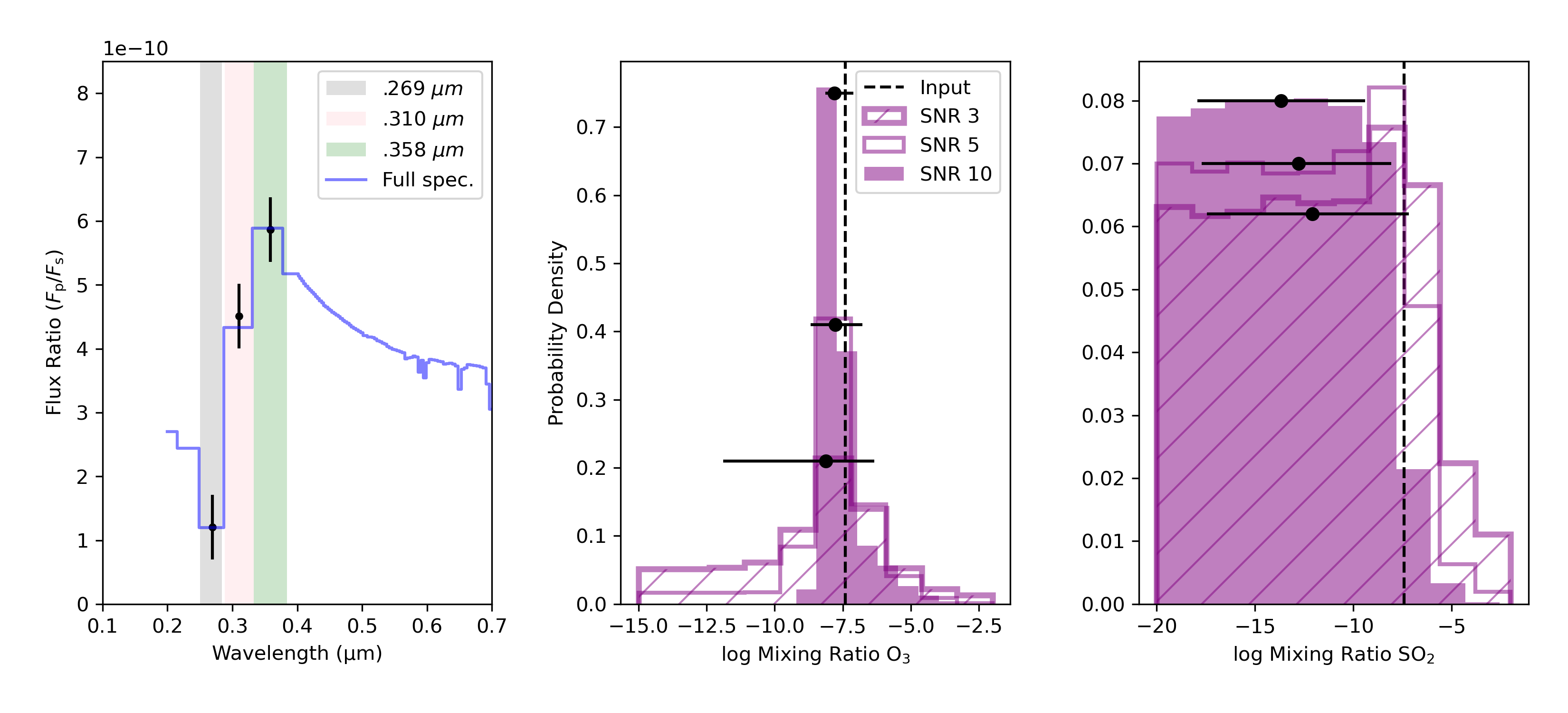}
\end{tabular}
\end{center}
\caption 
{ \label{DISRA_2B}
Retrieval results for the \textbf{2B} bandpass configuration under a high \ce{SO2} flux scenario with 0.1\% present day \ce{O2}. Left: Forward-model spectrum with simulated data with SNR=10 set relative to the flux at 0.33\,$\upmu$m shown for reference. Grey shading marks the 0.25\,--\,0.288\,$\upmu$m bandpass, pink shading marks the 0.288\,--\,0.333\,$\upmu$m bandpass, and green shading marks the 0.333\,--\,0.384\,$\upmu$m bandpass. Middle: Marginal posterior distributions for the \ce{O3} abundance at SNR=3 (hatched), 5 (unfilled), and 10 (filled). Right: Same as the middle panel but for the \ce{SO2} abundance. Horizontal bars above each posterior indicate the $\pm$1\,$\sigma$ regions, and the black point marks the mean value. Black vertical dashed lines indicate the truth (input) column-averaged mixing ratio abundance for each species.} 
\end{figure}

In Figure \ref{DISRA_2B}, three side-by-side 15\% bandpasses are positioned over the \ce{O3} feature probing the full extent of the spectral slope from 0.250\,--\,0.384\,$\upmu$m. The simulated photometric data is shown with errorbars corresponding to an SNR of 10 for reference. The middle panel displays the marginal posterior distributions for SNR values of 3, 5, and 10. For SNR 10, and SNR 5 the 1\,$\sigma$ spreads are approximately an order of magnitude, indicating tight constraints. The rightmost panel shows the marginal posterior for the \ce{SO2} abundance under the high flux scenario. \ce{SO2} remains largely unconstrained across all SNR cases with a potential upper limit constraint at SNR 10.


\begin{figure} [H]
\begin{center}
\begin{tabular}{c}
\includegraphics[height=7.5cm]{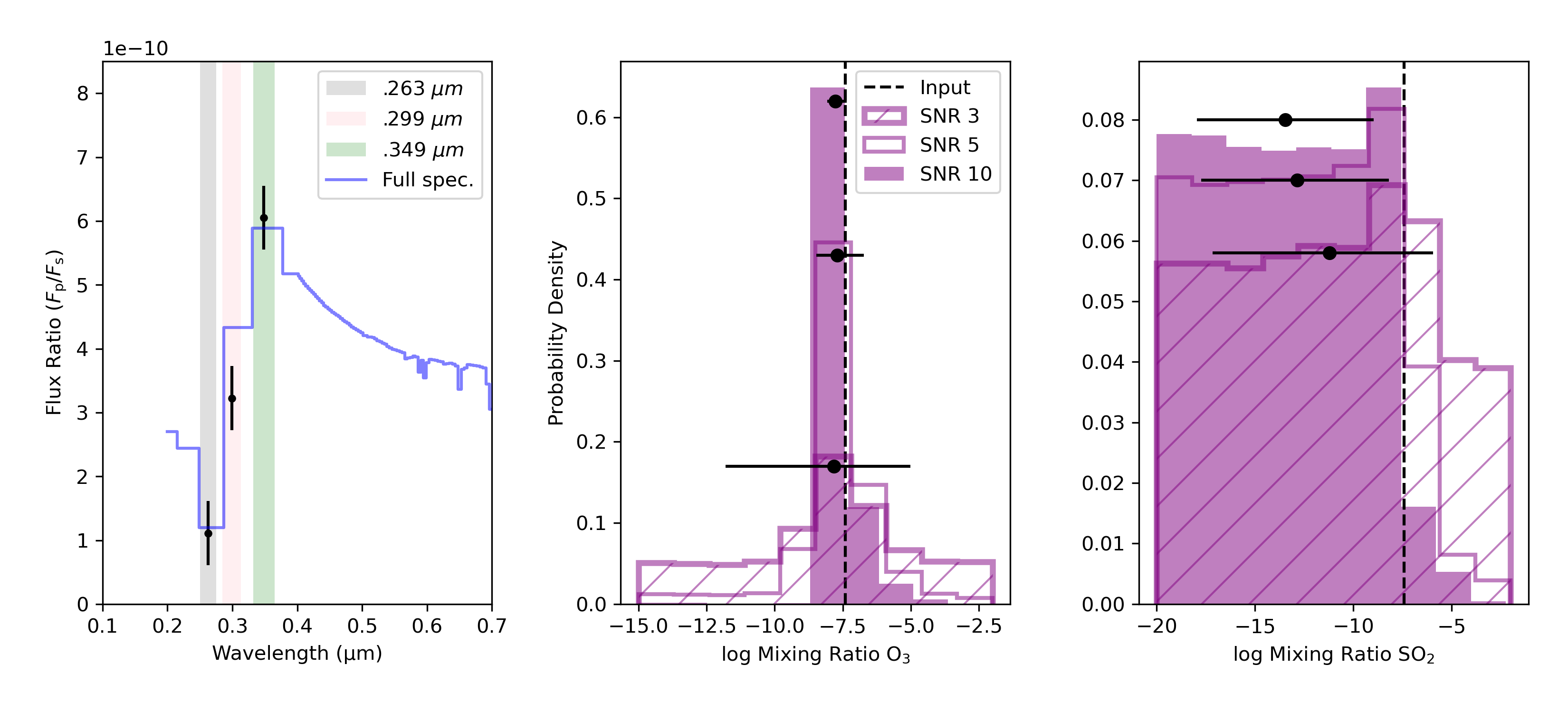}
\end{tabular}
\end{center}
\caption 
{ \label{DISRA_3A}
Retrieval results for the \textbf{3A} bandpass configuration under a high \ce{SO2} flux scenario with 0.1\% present day \ce{O2}. Left: Forward-model spectrum with simulated data with SNR=10 set relative to the flux at 0.33\,$\upmu$m shown for reference. Grey shading marks the 0.25\,--\,0.275\,$\upmu$m bandpass, pink shading marks the 0.285\,--\,0.313\,$\upmu$m bandpass, and green shading marks the 0.332\,--\,0.365\,$\upmu$m bandpass. Middle: Marginal posterior distributions for the \ce{O3} abundance at SNR=3 (hatched), 5 (unfilled), and 10 (filled). Right: Same as the middle panel but for the \ce{SO2} abundance. Horizontal bars above each posterior indicate the $\pm$1\,$\sigma$ regions, and the black point marks the mean value. Black vertical dashed lines indicate the truth (input) column-averaged mixing ratio abundance for each species. } 
\end{figure}

The 3A bandpass configuration spans 0.250\,--\,0.365\,$\upmu$m and includes three 10\% bandpasses separated by small gaps to ensure precise positioning. Similar to the 2B retrievals, tight constraints are able to be obtained at SNRs above 3 for \ce{O3} while the \ce{SO2} goes largely unconstrained for all cases (Figure \ref{DISRA_3A}). 


\begin{figure} [H]
\begin{center}
\begin{tabular}{c}
\includegraphics[height=7.5cm]{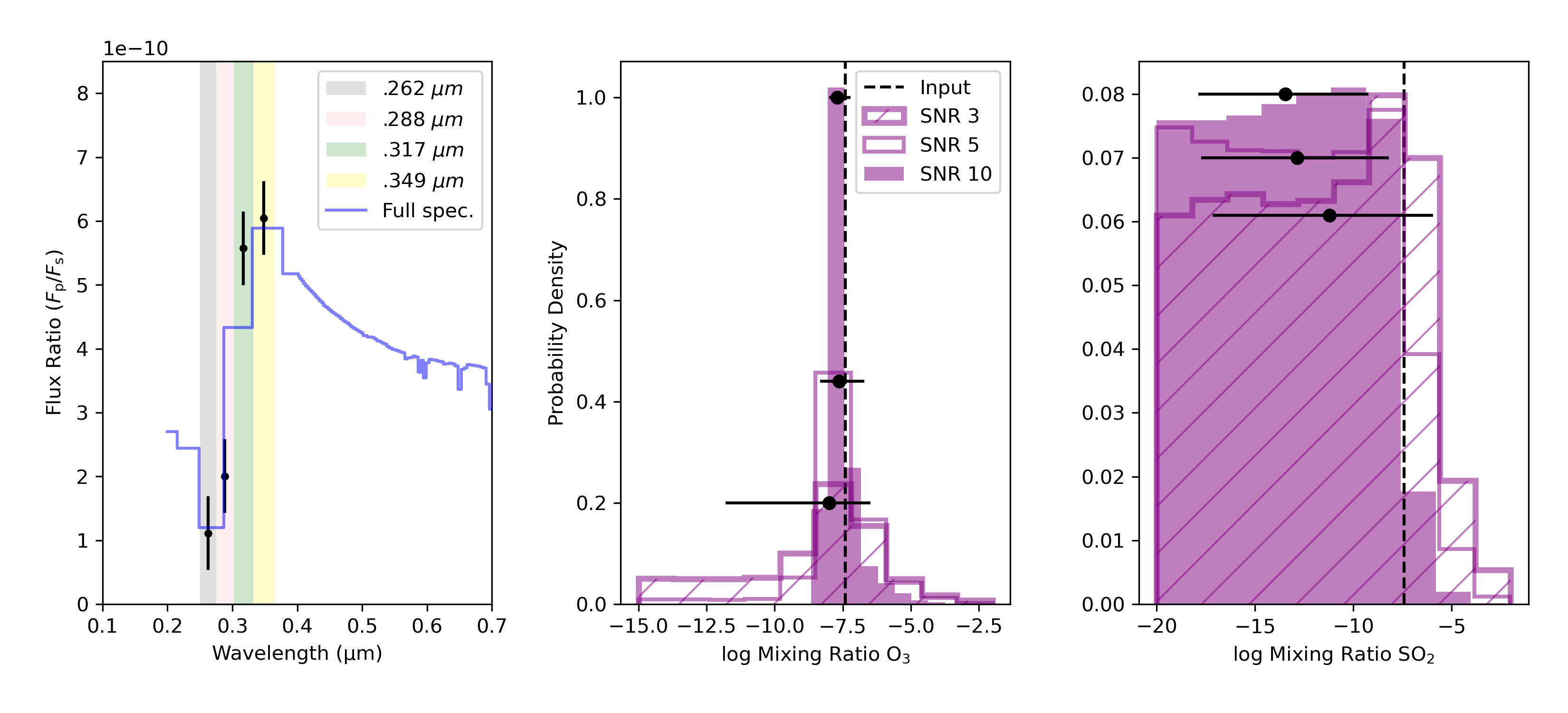}
\end{tabular}
\end{center}
\caption 
{ \label{DISRA_3B}
Retrieval results for the \textbf{3B} bandpass configuration under a high \ce{SO2} flux scenario with 0.1\% present day \ce{O2}. Left: Forward-model spectrum with simulated data with SNR=10 set relative to the flux at 0.33\,$\upmu$m shown for reference. Grey shading marks the 0.25\,--\,0.275\,$\upmu$m bandpass, pink shading marks the 0.275\,--\,0.302\,$\upmu$m bandpass, green shading marks the 0.302\,--\,0.332\,$\upmu$m bandpass, and yellow shading marks the 0.332\,--\,0.365\,$\upmu$m bandpass. Middle: Marginal posterior distributions for the \ce{O3} abundance at SNR=3 (hatched), 5 (unfilled), and 10 (filled). Right: Same as the middle panel but for the \ce{SO2} abundance. Horizontal bars above each posterior indicate the $\pm$1\,$\sigma$ regions, and the black point marks the mean value. Black vertical dashed lines indicate the truth (input) column-averaged mixing ratio abundance for each species. } 
\end{figure}

The 3B bandpass configuration (Figure \ref{DISRA_3B}) also spans 0.250\,--\,0.365\,$\upmu$m, similar to 3A, but consists of four adjacent 10\% bandpasses. As with 3A, tight constraints are achieved for \ce{O3} at SNRs 5 and 10, while \ce{SO2} remains unconstrained. The 1$\sigma$ uncertainties show marginal improvement compared to the 3A case.


\subsection{\ce{O3} Detection Significance}

We evaluated the strength of \ce{O3} detections in our modeling results by calculating the Bayesian Information Criterion (BIC) which evaluates model selection (in terms of the number of fitted parameters) against the goodness of fit between the model and simulated dataset. By performing retrieval simulations both with and without a given parameter of interest (in this case both \ce{O3} and \ce{O2}), taking the difference in BIC values ($\Delta$BIC) between the two retrieval cases can reveal whether the addition of that given parameter is favored over excluding it. In this case, \ce{O2} was removed in addition to \ce{O3} given the interdependency between these two species and allowed us to generate a representative ``null" case. The BIC is calculated using the following equation

\begin{equation}
    \rm{BIC} = n\ln{(N)} - 2\ln{\mathcal{(\hat{L})}}
\end{equation}
where n is the number of fitted parameters, N is the number of datapoints, and $\mathcal{\hat{L}}$ is the maximum likelihood \cite{schwarz_1978}. The preferred model setup is typically the one with a lower BIC value. The more negative the $\Delta$BIC, the stronger the preference is for including the additional free parameters of interest in the modeling setup. In terms of detection strength, we cross reference our $\Delta$BIC with previously established evidence criteria \cite{kass1995bayes,jeffreys1998theory} and find that, at SNR 10, configurations 2B, 3A, and 3B show strong evidence for the inclusion of \ce{O3} and \ce{O2}. These results would correspond to weak detections (or ``minor" results) in terms of Bayes Factor interpretations \cite{Thorngren_2026}. Table \ref{tab_BIC_Bayes} below includes the $\Delta$BIC and the corresponding level of evidence for each photometric bandpass configuration for the 0.1\% present level of \ce{O2} atmospheric scenario.

\begin{table}[H]
    \centering
    \begin{tabular}{c|c|c|c}
    \hline
    \hline
         Config. & SNR & $\Delta$BIC & Evidence \\
        \hline
        \hline
         
        1A  & 5 & 1.39 & Null hypothesis supported \\
            & 10 & 1.38 & ``\hspace{100pt}" \\
        \hline
        2B  & 5 & 0.05 & Null hypothesis supported \\
            & 10 & -6.38 & Strong Evidence \\
        \hline
        3A  & 5 & -0.15 & Not worth mentioning \\
            & 10 & -6.96 & Strong Evidence \\
        \hline
        3B  & 5 & 0.14 & Null hypothesis supported \\
            & 10 & -7.64 & Strong Evidence \\
        \hline  
    \end{tabular}
    \caption{Table summarizing the detection significance of \ce{O3} for each photometric bandpass retrieval scenario, reporting the corresponding $\Delta$BIC and associated level of evidence under the assumption of a 0.1\% present-day \ce{O2} Proterozoic atmospheric composition.}
    \label{tab_BIC_Bayes}
\end{table}

\section{Discussion} 
\label{sec:Disc}

Access to the NUV wavelength range on HWO is essential for interpreting and characterizing Earth-like planets in less oxygenated states. At visible wavelengths, \ce{O2} and \ce{O3} spectral features are weak and difficult to detect at low abundances, making NUV sensitivity critical for identifying biosignatures in such atmospheres. Since Earth spent nearly half its history in a weakly oxygenated state where \ce{O2} was present but substantially below modern levels, maintaining NUV capabilities is vital for ensuring that HWO can detect and study planets representative of this evolutionary phase. Without this sensitivity, the mission risks falling short of scientific goals outlined in the decadal survey. 

The retrieval study was conducted under idealized noise assumptions to isolate the fundamental detectability of \ce{O3} in the NUV. While spectra with randomized noise introduce additional scatter in the simulated spectra, their impact on the retrieved \ce{O3} abundance is expected to be modest compared to the uncertainties introduced by instrument-specific noise sources, throughput, variations, and detector performance. A full exploration of these effects will require coupling the retrieval framework to detailed instrument models, which is an area the authors aim to pursue in future work. In addition, while the number of free parameters in our retrievals exceeds the number of spectral data points, this does not pose a problem in a Bayesian framework. Parameters that are not informed by the data naturally return posterior distributions that reflect their priors, while parameters that influence the spectrum become constrained. This approach allows the retrieval to marginalize over nuisance parameters and remain agnostic about which parameter properties the data are sensitive to. Prior studies \cite{Line_2013} note that underdetermined parameter spaces are problematic primarily for frequentist techniques such as Optimal Estimation or bootstrap Monte Carlo, which require a unique, well constrained solution.

The retrieval results presented here establish the baseline observational requirements for detecting \ce{O3} in the NUV for Proterozoic Earth-like scenarios, while also addressing the risk of spectral false positives from \ce{SO2} absorption within this wavelength range. The shortest wavelength cutoff yielded the tightest constraints on the \ce{O3} abundance. However, pushing observations into the far-UV would introduce significant observational and technical challenges. For example, although the signal-to-noise is set relative to the planetary brightness at 0.36\,$\upmu$m, both the planetary and stellar fluxes drop sharply at shorter wavelengths. As a result, the absolute photon counts in these bins drastically decrease, which would result in a dramatic reduction in planetary yield. Importantly, the 1\,$\sigma$ constraints on the \ce{O3} abundance remain to within an order of magnitude out to approximately 0.26\,$\upmu$m making this a practical and sufficient observational cutoff for achieving strong \ce{O3} detections. This cutoff is also thought to be technologically feasible given that broadband photometry can be used in lieu of spectroscopy, but further laboratory experiments are necessary to further develop UV coronographic technology. The short wavelength cutoff of the instrument is recommended to extend down to 0.2\,$\upmu$m to mitigate edge-of-bandpass noise and optimize observational performance, particularly given the sensitivity and throughput challenges inherent to UV coronography. The long wavelength cutoff for observations should include coverage of the continuum and sensitivity to scattering opacity sources such as background gases and haze. Furthermore, ensuring that the long wavelength cutoff of the NUV instrument overlaps with the short-wavelength limit of the visible coronagraph would provide critical spectral continuity, allowing observations to be coherently inferred across instruments. In practice, although background and systematic noise will pose observational challenges to contend with, recent estimates indicate that the required integration times remain plausible (though demanding at $\sim 10$ days) subject to continued advances in NUV coronagraph instrument design \cite{Morgan_2024}. 

Our retrieval results demonstrate that robust constraints on the \ce{O3} abundance can be achieved through photometric observations using only a small number of strategically placed bandpasses. Ideally, four adjacent 10\% bandpasses would allow for fine sensitivity to the \ce{O3} spectral slope and offer maximal flexibility among the bandpass configurations tested here to tailor bandpass selection on a target-by-target basis. However, narrower bandpasses limit photon collection per bin, necessitating longer integration times to reach the desired SNR. Additionally, while the 3A bandpass configuration also produces strong constraints on the \ce{O3} abundance, a narrower bandpass width would still necessitate longer integration times for each observation. To balance spectral resolution with observational efficiency, we recommend a configuration of the three adjacent bandpasses (configuration 2B), each with a 15\% width. This setup still enables strong statistical evidence for \ce{O3} detections while potentially reducing the integration time per observation. In addition, our retrievals show that \ce{O3} can be tightly constrained at SNRs as low as 5, even in the presence of of \ce{SO2}, which does not impede the signal. While reducing the SNR requirement by a factor of two would translate to a fourfold decrease in the required integration time, it is noted that the statistical evidence for detection is drastically reduced in comparison to the SNR 10 results. Previous work showed that \ce{O3} can be detected at a moderate resolving power of 7 and reliably distinguished from \ce{SO2} \cite{Damiano_2023}. Here we extend this by demonstrating that \ce{O3} and \ce{SO2} remain distinguishable at lower spectral resolution using photometry in contrast to spectroscopy.

Our photochemical-climate modeling results suggest that it is difficult to build up appreciable amounts of \ce{SO2} in an atmosphere that also has a substantial abundance of water vapor. This is consistent with previous work, where the presence of an ocean limits atmospheric \ce{SO2} by acting as a sink, reducing its availability for upper-atmosphere detection \cite{Loftus_2019,Ostberg_2023}. For especially explosive volcanic eruptions, where \ce{SO2} can reach higher altitudes, \ce{SO2} can serve as a precursor species for haze production further limiting the buildup of appreciable amounts of \ce{SO2}. This is reinforced with the retrieval results for \ce{SO2} which show that even with high \ce{SO2} outgassing that produces \ce{SO2} surface abundances comparable to the surface abundance of \ce{O3}, \ce{SO2} goes largely unconstrained for the bandpass configurations studied here. Out of all of our simulations, the only configuration in which the retrieval produces weak constraints of both species is with the two adjacent bandpasses each with a 22\% bandpass width (configuration 1A) where this placement lacks sufficient sensitivity to the spectral slope making it difficult to distinguish between the \ce{O3} and \ce{SO2} features.

Although many molecules exhibit electronic transitions somewhere in the ultraviolet wavelength range, only a limited subset produce strong, structured absorption features in the NUV that could plausibly mimic \ce{O3}. UV photons readily excite or photolyze molecules, so absorption in the far UV is common across a wide range of species. However, NUV absorption is comparatively rare, and among known atmospheric molecules, \ce{O3}, \ce{SO2}, and \ce{H2S} are some known examples of dominant absorbers in the NUV. This work, and prior work has demonstrated that differences in the resulting spectral slope from these features is sufficient to distinguish between them at modest spectral resolution and SNR \cite{Damiano_2023}.

The possibility still remains of less-studied or ``exotic absorbers" with features in the far to NUV. Venus provides a well-known example where despite decades of study, the identity of its NUV absorber remains uncertain \cite{Titov_2018,marcq_2018}. A more comprehensive analysis of molecular UV cross-sections, particularly for species relevant to reducing, sulfur-rich, or photochemically unusual atmospheres, would help quantify the full landscape of atmospheric environments that could be characterized using NUV observations.

Importantly, spectral discrimination need not rely solely on the NUV bandpass for making inferences based on singular species detections. Many hypothetical atmospheric scenarios would be chemically incompatible with the environmental conditions required to sustain appreciable abundances of \ce{O3}. For example, species associated with highly reducing atmospheres harboring thick haze are unlikely to coexist with substantial \ce{O3}, because photochemical oxygen production requires low reducing power and haze formation is typically enhanced in \ce{CH4} rich, \ce{O2} poor environments. Consequently, environmental context like surface conditions, atmospheric composition, and redox state, provides an additional layer of discrimination that helps rule out exotic absorbers even when retrieval inferences of observations are degenerate with multiple chemical species detections.

Atmospheric false positives for generating \ce{O2} (and by extension \ce{O3}) have been identified in specific planetary and stellar environments, including \ce{CO2} photolysis with limited HOx recycling, water loss and H-escape, and in the context of M dwarf systems with potentially highly irradiative UV environments \cite{Wordsworth_2014,Harman_2015,Luger_2015,Meadows_2018}. In this study, we focus on a different regime where these Proterozoic Earth-like atmospheres are weakly oxygenated and fundamentally assume that \ce{O2} is biologically sourced at the 0.1\% and 1\% PAL, but the emphasis on evaluating species detections based on the atmospheric and environmental context is critical for robust exoplanet characterization strategies.

The \ce{O3} abundance and its relation to inferrable abundances of \ce{O2} vary drastically with stellar type because the photochemical mechanisms for \ce{O3} production are directly impacted by UV stellar flux of the host star \cite{Segura_2003,Rugheimer_2013,Kozakis_2022}. In terms of biosignature detection in general, K dwarfs offer the advantage of longer photochemical lifetimes for species like \ce{CH4} which has a tendency to be rapidly destroyed in oxygenated environments without substantial biological production rates of \ce{CH4} \cite{Arney_2019}. While \ce{CH4} does not have any substantial absorption features in the NUV, observations of spectrally strong absorption of \ce{O3} (and/or \ce{O2}) alongside the detection of \ce{CH4} at longer wavelengths, would provide stronger evidence for biosignature detection \cite{lovelock_physical_1965,Hitchcock_1967,Lippincott_1967,Lovelock1975}. Other spectral host types brighter than G dwarfs offer better planet-to-star contrast enabling higher SNR for a given exposure time. Extending these simulations to Earth-like planets around a broader range of sun-like hosts accessible to HWO would help map this trade space more fully and inform future instrument design and optimization.

\section{Conclusion} 
\label{sec:conclusion}
Planetary observations with an NUV coronagraph instrument are essential for characterizing Proterozoic Earth-like planets where direct detection of \ce{O2} may be infeasible and biosignature inference relies on robust detection of \ce{O3}. To confidently detect \ce{O3} at abundances as low as 0.1\% the present atmospheric level, our results recommend a short wavelength cutoff for \ce{O3} observations near 0.26\,$\upmu$m and photometric observations using three adjacent spectral bandpasses, each with a 15\% width. This configuration enables 1\,$\sigma$ constraints to within an order of magnitude, which is sufficient to distinguish \ce{O3} from potential spectral false positives such as \ce{SO2}. 
In our simulations, surface \ce{SO2} abundances resulting from 100$\times$ present-day volcanic fluxes still allowed for clear spectral separation between \ce{O3} and \ce{SO2} under moderate NUV coverage, thus highlighting the robustness of \ce{O3} detectability even under elevated outgassing scenarios. By optimizing spectral coverage and bandpass configuration, this approach ensures that HWO remains sensitive to key biosignatures across a range of planetary states, unlocking the potential to detect life on worlds that resemble earlier inhabited periods of Earth through time.

\subsection*{Disclosures}
The authors declare that there are no financial interests, commercial affiliations, or other potential conflicts of interest that could have influenced the objectivity of this research or the writing of this paper. 

\subsection* {Code, Data, and Materials Availability}
The following software used in this work can be found via the following links:\\ \texttt{rfast}: \url{https://github.com/hablabx/rfast},\\ \texttt{Atmos}: \url{https://github.com/VirtualPlanetaryLaboratory/atmos}.\\ The data associated with generating each figure can be obtained via: \textbf{Zenodo link, TBD}

\subsection* {Acknowledgments}
 A. Young would like to acknowledge support from center-directed funding through a NASA Scientific Task Group. A. Young would like to acknowledge support from the GSFC Sellers Exoplanet Environments Collaboration (SEEC), which is supported by NASA's Planetary Science Division's Research Program. A. Young would also like to acknowledge support from the NASA Pathways Internship Program, which provided training and resources contributing to this research. R. Juanola-Parramon was supported by NASA through the CRESST II cooperative agreement CA 80GSFC24M0006. Resources supporting this work were provided by the NASA High-End Computing (HEC) Program through the NASA Advanced Supercomputing (NAS) Division at Ames Research Center. Portions of this text were refined with the assistance of an AI language model (Microsoft Copilot), which was used for editing and phrasing. All scientific content, analysis, and interpretation were developed solely by the authors. Part of this research was carried out at the Jet Propulsion Laboratory, California Institute of Technology, under a contract with the National Aeronautics and Space Administration (80NM0018D0004).

\appendix    
\section{Full Retrieval Example}
\begin{figure} [H]
\begin{center}
\begin{tabular}{c}
\includegraphics[height=15.9cm]{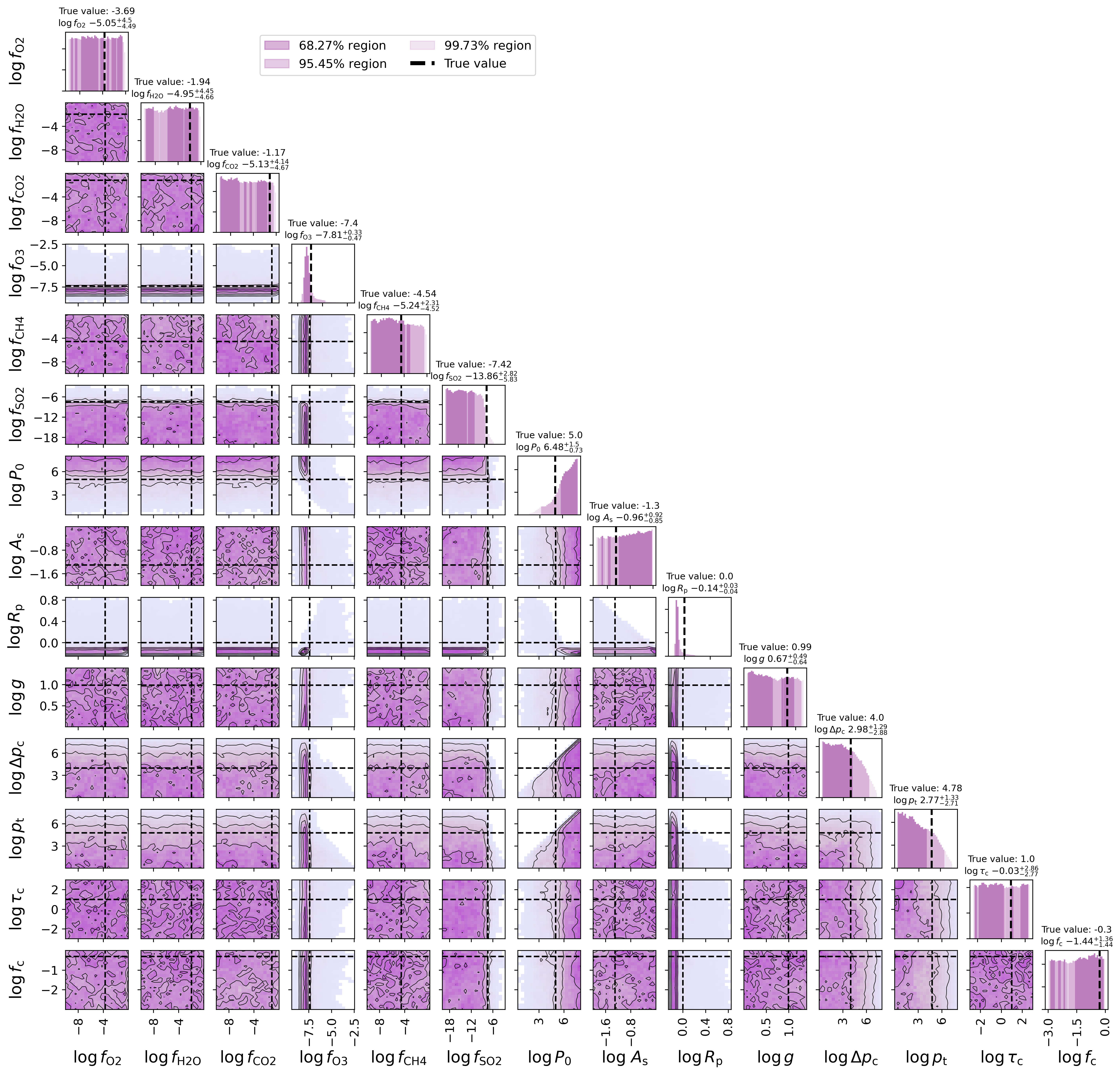}
\end{tabular}
\end{center}
\caption 
{ \label{DISRA_2B_Full}
The full 14-parameter corner plot for the \textbf{2B} bandpass configuration at SNR \textcolor{purple}{10}. One-dimensional marginal posterior distributions are shown along the diagonal with associated truth values and the 16$^{\rm th}$, 50$^{\rm th}$, and 84$^{\rm th}$ percentile values indicated above each marginal distribution. The truth values for each parameter are also indicated by black dashed lines.} 
\end{figure}

Figure \ref{DISRA_2B_Full} depicts the full retrieval results for the \textbf{2B} recommended bandpass configuration at SNR 10 described in Section \ref{sec:results}. Above each 1-D posterior, the truth values, and 1\,$\sigma$ credible regions are reported. The retrieval framework employed 14 parameters, many of which remained unconstrained due to the limited spectral information. However, the lack of constraints on these ancillary parameters did not compromise the robustness of the \ce{O3} detections.

\bibliography{references}   

@article{Lyons_2014,
  title={The rise of oxygen in Earth’s early ocean and atmosphere},
  author={Lyons, Timothy W and Reinhard, Christopher T and Planavsky, Noah J},
  journal={Nature},
  volume={506},
  number={7488},
  pages={307--315},
  year={2014},
  publisher={Nature Publishing Group UK London}
}

@ARTICLE{Planavsky_2014,
       author = {{Planavsky}, Noah J. and {Reinhard}, Christopher T. and {Wang}, Xiangli and {Thomson}, Danielle and {McGoldrick}, Peter and {Rainbird}, Robert H. and {Johnson}, Thomas and {Fischer}, Woodward W. and {Lyons}, Timothy W.},
        title = "{Low Mid-Proterozoic atmospheric oxygen levels and the delayed rise of animals}",
      journal = {Science},
         year = 2014,
        month = oct,
       volume = {346},
       number = {6209},
        pages = {635-638},
          doi = {10.1126/science.1258410},
       adsurl = {https://ui.adsabs.harvard.edu/abs/2014Sci...346..635P}
}

@ARTICLE{Reinhard_2017,
       author = {{Reinhard}, Christopher T. and {Olson}, Stephanie L. and {Schwieterman}, Edward W. and {Lyons}, Timothy W.},
        title = "{False Negatives for Remote Life Detection on Ocean-Bearing Planets: Lessons from the Early Earth}",
      journal = {Astrobiology},
         year = 2017,
        month = apr,
       volume = {17},
       number = {4},
        pages = {287-297},
          doi = {10.1089/ast.2016.1598},
archivePrefix = {arXiv},
       eprint = {1702.01137},
 primaryClass = {astro-ph.EP},
       adsurl = {https://ui.adsabs.harvard.edu/abs/2017AsBio..17..287R}
}

@ARTICLE{Schwieterman_2018,
       author = {{Schwieterman}, Edward W. and {Kiang}, Nancy Y. and {Parenteau}, Mary N. and {Harman}, Chester E. and {DasSarma}, Shiladitya and {Fisher}, Theresa M. and {Arney}, Giada N. and {Hartnett}, Hilairy E. and {Reinhard}, Christopher T. and {Olson}, Stephanie L. and {Meadows}, Victoria S. and {Cockell}, Charles S. and {Walker}, Sara I. and {Grenfell}, John Lee and {Hegde}, Siddharth and {Rugheimer}, Sarah and {Hu}, Renyu and {Lyons}, Timothy W.},
        title = "{Exoplanet Biosignatures: A Review of Remotely Detectable Signs of Life}",
      journal = {Astrobiology},
         year = 2018,
        month = jun,
       volume = {18},
       number = {6},
        pages = {663-708},
          doi = {10.1089/ast.2017.1729},
archivePrefix = {arXiv},
       eprint = {1705.05791},
 primaryClass = {astro-ph.EP},
       adsurl = {https://ui.adsabs.harvard.edu/abs/2018AsBio..18..663S}
}

@ARTICLE{Pavlov_2003,
       author = {{Pavlov}, Alexander A. and {Hurtgen}, Matthew T. and {Kasting}, James F. and {Arthur}, Michael A.},
        title = "{Methane-rich Proterozoic atmosphere?}",
      journal = {Geology},
         year = 2003,
        month = jan,
       volume = {31},
       number = {1},
        pages = {87},
          doi = {10.1130/0091-7613(2003)031<0087:MRPA>2.0.CO;2},
       adsurl = {https://ui.adsabs.harvard.edu/abs/2003Geo....31...87P}
}

@ARTICLE{Claire_2012,
       author = {{Claire}, Mark W. and {Sheets}, John and {Cohen}, Martin and {Ribas}, Ignasi and {Meadows}, Victoria S. and {Catling}, David C.},
        title = "{The Evolution of Solar Flux from 0.1 nm to 160 {\ensuremath{\mu}}m: Quantitative Estimates for Planetary Studies}",
      journal = {The Astrophysical Journal},
         year = 2012,
        month = sep,
       volume = {757},
       number = {1},
          eid = {95},
        pages = {95},
          doi = {10.1088/0004-637X/757/1/95},
       adsurl = {https://ui.adsabs.harvard.edu/abs/2012ApJ...757...95C}
}

@ARTICLE{Zahnle_2006,
       author = {{Zahnle}, K. and {Claire}, M. and {Catling}, D.},
        title = "{The loss of mass-independent fractionation in sulfur due to a Palaeoproterozoic collapse of atmospheric methane}",
      journal = {Geobiology},
         year = 2006,
        month = dec,
       volume = {4},
       number = {4},
        pages = {271-283},
          doi = {10.1111/j.1472-4669.2006.00085.x},
       adsurl = {https://ui.adsabs.harvard.edu/abs/2006Gbio....4..271Z}
}

@ARTICLE{Robinson_2023,
       author = {{Robinson}, Tyler D. and {Salvador}, Arnaud},
        title = "{Exploring and Validating Exoplanet Atmospheric Retrievals with Solar System Analog Observations}",
      journal = {The Planetary Science Journal},
         year = 2023,
        month = jan,
       volume = {4},
       number = {1},
          eid = {10},
        pages = {10},
          doi = {10.3847/PSJ/acac9a},
archivePrefix = {arXiv},
       eprint = {2204.04231},
 primaryClass = {astro-ph.EP},
       adsurl = {https://ui.adsabs.harvard.edu/abs/2023PSJ.....4...10R}
}

@ARTICLE{Teal_2022,
       author = {{Teal}, D.~J. and {Kempton}, Eliza M. -R. and {Bastelberger}, Sandra and {Youngblood}, Allison and {Arney}, Giada},
        title = "{Effects of UV Stellar Spectral Uncertainty on the Chemistry of Terrestrial Atmospheres}",
      journal = {The Astrophysical Journal},
         year = 2022,
        month = mar,
       volume = {927},
       number = {1},
          eid = {90},
        pages = {90},
          doi = {10.3847/1538-4357/ac4d99},
archivePrefix = {arXiv},
       eprint = {2201.08805},
 primaryClass = {astro-ph.EP},
       adsurl = {https://ui.adsabs.harvard.edu/abs/2022ApJ...927...90T}
}

@ARTICLE{Loftus_2019,
       author = {{Loftus}, Kaitlyn and {Wordsworth}, Robin D. and {Morley}, Caroline V.},
        title = "{Sulfate Aerosol Hazes and SO$_{2}$ Gas as Constraints on Rocky Exoplanets{\textquoteright} Surface Liquid Water}",
      journal = {Astrophysical Journal},
         year = 2019,
        month = dec,
       volume = {887},
       number = {2},
          eid = {231},
        pages = {231},
          doi = {10.3847/1538-4357/ab58cc},
archivePrefix = {arXiv},
       eprint = {1908.02769},
 primaryClass = {astro-ph.EP},
       adsurl = {https://ui.adsabs.harvard.edu/abs/2019ApJ...887..231L}
}

@ARTICLE{Damiano_2023,
       author = {{Damiano}, Mario and {Hu}, Renyu and {Mennesson}, Bertrand},
        title = "{Reflected Spectroscopy of Small Exoplanets. III. Probing the UV Band to Measure Biosignature Gases}",
      journal = {The Astronomical Journal},
         year = 2023,
        month = oct,
       volume = {166},
       number = {4},
          eid = {157},
        pages = {157},
          doi = {10.3847/1538-3881/acefd3},
archivePrefix = {arXiv},
       eprint = {2308.08490},
 primaryClass = {astro-ph.EP},
       adsurl = {https://ui.adsabs.harvard.edu/abs/2023AJ....166..157D}
}

@ARTICLE{Foreman_Mackey_2013,
       author = {{Foreman-Mackey}, Daniel and {Hogg}, David W. and {Lang}, Dustin and
         {Goodman}, Jonathan},
        title = "{emcee: The MCMC Hammer}",
      journal = {Publications of the Astronomical Society of the Pacific},
         year = 2013,
        month = mar,
       volume = {125},
       number = {925},
        pages = {306},
          doi = {10.1086/670067},
archivePrefix = {arXiv},
       eprint = {1202.3665},
 primaryClass = {astro-ph.IM},
       adsurl = {https://ui.adsabs.harvard.edu/abs/2013PASP..125..306F}
}

@ARTICLE{Mennesson_2024,
       author = {{Mennesson}, Bertrand and {Belikov}, Ruslan and {Por}, Emiel and {Serabyn}, Eugene and {Ruane}, Garreth and {Riggs}, A.~J. Eldorado and {Sirbu}, Dan and {Pueyo}, Laurent and {Soummer}, Remi and {Kasdin}, Jeremy and {Shaklan}, Stuart and {Seo}, Byoung-Joon and {Stark}, Christopher and {Cady}, Eric and {Chen}, Pin and {Crill}, Brendan and {Fogarty}, Kevin and {Greenbaum}, Alexandra and {Guyon}, Olivier and {Juanola-Parramon}, Roser and {Kern}, Brian and {Krist}, John and {Macintosh}, Bruce and {Marx}, David and {Mawet}, Dimitri and {Prada}, Camilo Mejia and {Morgan}, Rhonda and {Nemati}, Bijan and {Pogorelyuk}, Leonid and {Redmond}, Susan and {Seager}, Sara and {Siegler}, Nicholas and {Stapelfeldt}, Karl and {Steiger}, Sarah and {Trauger}, John and {Wallace}, James K. and {Ygouf}, Marie and {Zimmerman}, Neil},
        title = "{Current laboratory performance of starlight suppression systems and potential pathways to desired Habitable Worlds Observatory exoplanet science capabilities}",
      journal = {Journal of Astronomical Telescopes, Instruments, and Systems},
         year = 2024,
        month = jul,
       volume = {10},
          eid = {035004},
        pages = {035004},
          doi = {10.1117/1.JATIS.10.3.035004},
archivePrefix = {arXiv},
       eprint = {2404.18036},
 primaryClass = {astro-ph.IM},
       adsurl = {https://ui.adsabs.harvard.edu/abs/2024JATIS..10c5004M}
}

@INPROCEEDINGS{Morgan_2024,
       author = {{Morgan}, Rhonda and {Savransky}, Dmitry and {Turmon}, Michael and {Damiano}, Mario and {Hu}, Renyu and {Mennesson}, Bertrand and {Mamajek}, Eric E. and {Robinson}, Tyler D. and {Tokadjian}, Armen},
        title = "{HWO yield sensitivities in the NIR and NUV}",
    booktitle = {Space Telescopes and Instrumentation 2024: Optical, Infrared, and Millimeter Wave},
         year = 2024,
       editor = {{Coyle}, Laura E. and {Matsuura}, Shuji and {Perrin}, Marshall D.},
       series = {Society of Photo-Optical Instrumentation Engineers (SPIE) Conference Series},
       volume = {13092},
        month = aug,
          eid = {130925M},
        pages = {130925M},
          doi = {10.1117/12.3020858},
archivePrefix = {arXiv},
       eprint = {2410.16492},
 primaryClass = {astro-ph.EP},
       adsurl = {https://ui.adsabs.harvard.edu/abs/2024SPIE13092E..5MM}
}

@ARTICLE{Madhusudhan_2009,
       author = {{Madhusudhan}, N. and {Seager}, S.},
        title = "{A Temperature and Abundance Retrieval Method for Exoplanet Atmospheres}",
      journal = {The Astrophysical Journal},
         year = 2009,
        month = dec,
       volume = {707},
       number = {1},
        pages = {24-39},
          doi = {10.1088/0004-637X/707/1/24},
archivePrefix = {arXiv},
       eprint = {0910.1347},
 primaryClass = {astro-ph.EP},
       adsurl = {https://ui.adsabs.harvard.edu/abs/2009ApJ...707...24M}
}

@ARTICLE{Benneke_2012,
       author = {{Benneke}, Bjoern and {Seager}, Sara},
        title = "{Atmospheric Retrieval for Super-Earths: Uniquely Constraining the Atmospheric Composition with Transmission Spectroscopy}",
      journal = {Astrophysical Journal},
         year = 2012,
        month = jul,
       volume = {753},
       number = {2},
          eid = {100},
        pages = {100},
          doi = {10.1088/0004-637X/753/2/100},
archivePrefix = {arXiv},
       eprint = {1203.4018},
 primaryClass = {astro-ph.EP},
       adsurl = {https://ui.adsabs.harvard.edu/abs/2012ApJ...753..100B}
}

@ARTICLE{Line_2013,
       author = {{Line}, Michael R. and {Wolf}, Aaron S. and {Zhang}, Xi and {Knutson}, Heather and {Kammer}, Joshua A. and {Ellison}, Elias and {Deroo}, Pieter and {Crisp}, Dave and {Yung}, Yuk L.},
        title = "{A Systematic Retrieval Analysis of Secondary Eclipse Spectra. I. A Comparison of Atmospheric Retrieval Techniques}",
      journal = {Astrophysical Journal},
         year = 2013,
        month = oct,
       volume = {775},
       number = {2},
          eid = {137},
        pages = {137},
          doi = {10.1088/0004-637X/775/2/137},
archivePrefix = {arXiv},
       eprint = {1304.5561},
 primaryClass = {astro-ph.EP},
       adsurl = {https://ui.adsabs.harvard.edu/abs/2013ApJ...775..137L}
}

@ARTICLE{Lupu_2016,
       author = {{Lupu}, Roxana E. and {Marley}, Mark S. and {Lewis}, Nikole and {Line}, Michael and {Traub}, Wesley A. and {Zahnle}, Kevin},
        title = "{Developing Atmospheric Retrieval Methods for Direct Imaging Spectroscopy of Gas Giants in Reflected Light. I. Methane Abundances and Basic Cloud Properties}",
      journal = {Astronomical Journal},
         year = 2016,
        month = dec,
       volume = {152},
       number = {6},
          eid = {217},
        pages = {217},
          doi = {10.3847/0004-6256/152/6/217},
archivePrefix = {arXiv},
       eprint = {1604.05370},
 primaryClass = {astro-ph.IM},
       adsurl = {https://ui.adsabs.harvard.edu/abs/2016AJ....152..217L}
}

@ARTICLE{Feng_2018,
       author = {{Feng}, Y. Katherina and {Robinson}, Tyler D. and {Fortney}, Jonathan J. and {Lupu}, Roxana E. and {Marley}, Mark S. and {Lewis}, Nikole K. and {Macintosh}, Bruce and {Line}, Michael R.},
        title = "{Characterizing Earth Analogs in Reflected Light: Atmospheric Retrieval Studies for Future Space Telescopes}",
      journal = {Astronomical Journal},
         year = 2018,
        month = may,
       volume = {155},
       number = {5},
          eid = {200},
        pages = {200},
          doi = {10.3847/1538-3881/aab95c},
archivePrefix = {arXiv},
       eprint = {1803.06403},
 primaryClass = {astro-ph.EP},
       adsurl = {https://ui.adsabs.harvard.edu/abs/2018AJ....155..200F}
}

@ARTICLE{Barstow_2020,
       author = {{Barstow}, Joanna K. and {Changeat}, Quentin and {Garland}, Ryan and {Line}, Michael R. and {Rocchetto}, Marco and {Waldmann}, Ingo P.},
        title = "{A comparison of exoplanet spectroscopic retrieval tools}",
      journal = {Monthly Notices of the RAS},
         year = 2020,
        month = apr,
       volume = {493},
       number = {4},
        pages = {4884-4909},
          doi = {10.1093/mnras/staa548},
archivePrefix = {arXiv},
       eprint = {2002.01063},
 primaryClass = {astro-ph.EP},
       adsurl = {https://ui.adsabs.harvard.edu/abs/2020MNRAS.493.4884B}
}

@ARTICLE{MacDonald_2023,
       author = {{MacDonald}, Ryan J. and {Batalha}, Natasha E.},
        title = "{A Catalog of Exoplanet Atmospheric Retrieval Codes}",
      journal = {Research Notes of the American Astronomical Society},
         year = 2023,
        month = mar,
       volume = {7},
       number = {3},
          eid = {54},
        pages = {54},
          doi = {10.3847/2515-5172/acc46a},
archivePrefix = {arXiv},
       eprint = {2303.12925},
 primaryClass = {astro-ph.EP},
       adsurl = {https://ui.adsabs.harvard.edu/abs/2023RNAAS...7...54M}
}

@BOOK{NAS_2021,  
    author    ={{National Academies of Sciences, Engineering, and Medicine}},  title     = "Pathways to Discovery in Astronomy and Astrophysics for the 2020s",  
    isbn      = "978-0-309-46586-1",  
    doi       = "10.17226/26141",  
    url       = "https://nap.nationalacademies.org/catalog/26141/pathways-to-discovery-in-astronomy-and-astrophysics-for-the-2020s",
    year      =2021,
    publisher = "The National Academies Press",
    address   = "Washington, DC"
}

@ARTICLE{HabEx_Paper_2018,
       author = {{Gaudi}, B. Scott and {Seager}, Sara and {Mennesson}, Bertrand and {Kiessling}, Alina and {Warfield}, Keith R. and {Habitable Exoplanet Observatory Science} and {Technology Definition Team}},
        title = "{The Habitable Exoplanet Observatory}",
      journal = {Nature Astronomy},
         year = 2018,
        month = aug,
       volume = {2},
        pages = {600-604},
          doi = {10.1038/s41550-018-0549-2},
       adsurl = {https://ui.adsabs.harvard.edu/abs/2018NatAs...2..600G}
}

@ARTICLE{LUVOIR_Paper_2018,
       author = {{Roberge}, Aki and {Moustakas}, Leonidas A.},
        title = "{The Large Ultraviolet/Optical/Infrared Surveyor}",
      journal = {Nature Astronomy},
         year = 2018,
        month = aug,
       volume = {2},
        pages = {605-607},
          doi = {10.1038/s41550-018-0543-8},
       adsurl = {https://ui.adsabs.harvard.edu/abs/2018NatAs...2..605R}
}

@INPROCEEDINGS{Bailey_RST_2023,
       author = {{Bailey}, Vanessa P. and {Bendek}, Eduardo and {Monacelli}, Brian and {Baker}, Caleb and {Bedrosian}, Gasia and {Cady}, Eric and {Douglas}, Ewan S. and {Groff}, Tyler and {Hildebrandt}, Sergi R. and {Kasdin}, N. Jeremy and {Krist}, John and {Macintosh}, Bruce and {Mennesson}, Bertrand and {Morrissey}, Patrick and {Poberezhskiy}, Ilya and {Subedi}, Hari B. and {Rhodes}, Jason and {Roberge}, Aki and {Ygouf}, Marie and {Zellem}, Robert T. and {Zhao}, Feng and {Zimmerman}, Neil T.},
        title = "{Nancy Grace Roman Space Telescope coronagraph instrument overview and status}",
    booktitle = {Society of Photo-Optical Instrumentation Engineers (SPIE) Conference Series},
         year = 2023,
       series = {Society of Photo-Optical Instrumentation Engineers (SPIE) Conference Series},
       volume = {12680},
        month = oct,
          eid = {126800T},
        pages = {126800T},
          doi = {10.1117/12.2679036},
archivePrefix = {arXiv},
       eprint = {2309.08672},
 primaryClass = {astro-ph.IM},
       adsurl = {https://ui.adsabs.harvard.edu/abs/2023SPIE12680E..0TB}
}

@article{arney_pale_2016,
	title = {The {Pale} {Orange} {Dot}: {The} {Spectrum} and {Habitability} of {Hazy} {Archean} {Earth}},
	volume = {16},
	issn = {1531-1074},
	shorttitle = {The {Pale} {Orange} {Dot}},
	url = {https://www.liebertpub.com/doi/abs/10.1089/ast.2015.1422},
	doi = {10.1089/ast.2015.1422},
	number = {11},
	journal = {Astrobiology},
	author = {Arney, Giada and Domagal-Goldman, Shawn D. and Meadows, Victoria S. and Wolf, Eric T. and Schwieterman, Edward and Charnay, Benjamin and Claire, Mark and Hébrard, Eric and Trainer, Melissa G.},
	month = oct,
	year = {2016},
	pages = {873--899}
}

@ARTICLE{Arney_pale_2017,
       author = {{Arney}, Giada N. and {Meadows}, Victoria S. and {Domagal-Goldman}, Shawn D. and {Deming}, Drake and {Robinson}, Tyler D. and {Tovar}, Guadalupe and {Wolf}, Eric T. and {Schwieterman}, Edward},
        title = "{Pale Orange Dots: The Impact of Organic Haze on the Habitability and Detectability of Earthlike Exoplanets}",
      journal = {The Astrophysical Journal},
         year = 2017,
        month = feb,
       volume = {836},
       number = {1},
          eid = {49},
        pages = {49},
          doi = {10.3847/1538-4357/836/1/49},
archivePrefix = {arXiv},
       eprint = {1702.02994},
 primaryClass = {astro-ph.EP},
       adsurl = {https://ui.adsabs.harvard.edu/abs/2017ApJ...836...49A}
}

@ARTICLE{Livengood_2011,
       author = {{Livengood}, Timothy A. and {Deming}, L. Drake and {A'Hearn}, Michael F. and {Charbonneau}, David and {Hewagama}, Tilak and {Lisse}, Carey M. and {McFadden}, Lucy A. and {Meadows}, Victoria S. and {Robinson}, Tyler D. and {Seager}, Sara and {Wellnitz}, Dennis D.},
        title = "{Properties of an Earth-Like Planet Orbiting a Sun-Like Star: Earth Observed by the EPOXI Mission}",
      journal = {Astrobiology},
         year = 2011,
        month = nov,
       volume = {11},
       number = {9},
        pages = {907-930},
          doi = {10.1089/ast.2011.0614},
       adsurl = {https://ui.adsabs.harvard.edu/abs/2011AsBio..11..907L}
}

@ARTICLE{Arney_2019,
       author = {{Arney}, Giada N.},
        title = "{The K Dwarf Advantage for Biosignatures on Directly Imaged Exoplanets}",
      journal = {The Astrophysical Journal Letters},
         year = 2019,
        month = mar,
       volume = {873},
       number = {1},
          eid = {L7},
        pages = {L7},
          doi = {10.3847/2041-8213/ab0651},
archivePrefix = {arXiv},
       eprint = {2001.10458},
 primaryClass = {astro-ph.EP},
       adsurl = {https://ui.adsabs.harvard.edu/abs/2019ApJ...873L...7A}
}

@ARTICLE{Robinson_2014,
       author = {{Robinson}, Tyler D. and {Maltagliati}, Luca and {Marley}, Mark S. and {Fortney}, Jonathan J.},
        title = "{Titan solar occultation observations reveal transit spectra of a hazy world}",
      journal = {Proceedings of the National Academy of Science},
         year = 2014,
        month = jun,
       volume = {111},
       number = {25},
        pages = {9042-9047},
          doi = {10.1073/pnas.1403473111},
archivePrefix = {arXiv},
       eprint = {1406.3314},
 primaryClass = {astro-ph.EP},
       adsurl = {https://ui.adsabs.harvard.edu/abs/2014PNAS..111.9042R}
}

@ARTICLE{Bellucci_2009,
       author = {{Bellucci}, A. and {Sicardy}, B. and {Drossart}, P. and {Rannou}, P. and {Nicholson}, P.~D. and {Hedman}, M. and {Baines}, K.~H. and {Burrati}, B.},
        title = "{Titan solar occultation observed by Cassini/VIMS: Gas absorption and constraints on aerosol composition}",
      journal = {Icarus},
         year = 2009,
        month = may,
       volume = {201},
       number = {1},
        pages = {198-216},
          doi = {10.1016/j.icarus.2008.12.024},
       adsurl = {https://ui.adsabs.harvard.edu/abs/2009Icar..201..198B}
}

@ARTICLE{Hayne_2014,
       author = {{Hayne}, Paul O. and {McCord}, Thomas B. and {Sotin}, Christophe},
        title = "{Titan's surface composition and atmospheric transmission with solar occultation measurements by Cassini VIMS}",
      journal = {Icarus},
         year = 2014,
        month = nov,
       volume = {243},
        pages = {158-172},
          doi = {10.1016/j.icarus.2014.08.045},
       adsurl = {https://ui.adsabs.harvard.edu/abs/2014Icar..243..158H}
}

@ARTICLE{Maltagliati_2015,
       author = {{Maltagliati}, Luca and {B{\'e}zard}, Bruno and {Vinatier}, Sandrine and {Hedman}, Matthew M. and {Lellouch}, Emmanuel and {Nicholson}, Philip D. and {Sotin}, Christophe and {de Kok}, Remco J. and {Sicardy}, Bruno},
        title = "{Titan's atmosphere as observed by Cassini/VIMS solar occultations: CH$_{4}$, CO and evidence for C$_{2}$H$_{6}$ absorption}",
      journal = {Icarus},
         year = 2015,
        month = mar,
       volume = {248},
        pages = {1-24},
          doi = {10.1016/j.icarus.2014.10.004},
archivePrefix = {arXiv},
       eprint = {1405.6324},
 primaryClass = {astro-ph.EP},
       adsurl = {https://ui.adsabs.harvard.edu/abs/2015Icar..248....1M}
}

@ARTICLE{Meadows_1996,
       author = {{Meadows}, V.~S. and {Crisp}, D.},
        title = "{Ground-based near-infrared observations of the Venus nightside: The thermal structure and water abundance near the surface}",
      journal = {Journal of Goephysical Research},
         year = 1996,
        month = jan,
       volume = {101},
       number = {E2},
        pages = {4595-4622},
          doi = {10.1029/95JE03567},
       adsurl = {https://ui.adsabs.harvard.edu/abs/1996JGR...101.4595M}
}

@inproceedings{Young_2026,
  author    = {Young, Amber V. and Arney, Giada N. and Robinson, Tyler D. and Domagal-Goldman, Shawn D. and Juanola-Parramon, Roser and Stark, Christopher C. and Hu, Renyu and Tokadjian, Armen and Damiano, Mario},
  title     = {{Characterizing Proterozoic Earth-like Planets in the Near Ultraviolet with HWO}},
  booktitle = {Proceedings of Towards Habitable Worlds Observatory: Visionary Science and Transformational Technology},
  year      = {2026},
  note      = {In press}
}

@article{Gorkom_etal_2025,
author = {Kyle J. Van Gorkom and Ramya M. Anche and Christopher B. Mendillo and Jessica A. Gersh-Range and Justin Hom and Tyler D. Robinson and Mamadou N'Diaye and Nikole K. Lewis and Bruce A. Macintosh and Ewan S. Douglas},
title = {{Performance predictions and contrast limits for an ultraviolet high-contrast imaging testbed}},
volume = {11},
journal = {Journal of Astronomical Telescopes, Instruments, and Systems},
number = {4},
publisher = {SPIE},
pages = {042203},
year = {2025},
doi = {10.1117/1.JATIS.11.4.042203},
URL = {https://doi.org/10.1117/1.JATIS.11.4.042203}
}

@inproceedings{pueyo2019luvoir,
  title={The LUVOIR Extreme Coronagraph for Living Planetary Systems (ECLIPS) I: searching and characterizing exoplanetary gems},
  author={Pueyo, L and Stark, C and Juanola-Parramon, R and Zimmerman, N and Bolcar, M and Roberge, A and Arney, G and Ruane, G and Riggs, AJ and Belikov, R and others},
  booktitle={Techniques and Instrumentation for Detection of Exoplanets IX},
  volume={11117},
  pages={37--65},
  year={2019},
  organization={SPIE}
}

@inproceedings{juanola2019luvoir,
  title={The LUVOIR Extreme Coronagraph for Living Planetary Systems (ECLIPS) II. Performance evaluation, aberration sensitivity analysis and exoplanet detection simulations},
  author={Juanola-Parramon, Roser and Zimmerman, Neil T and Pueyo, Laurent and Bolcar, Matthew and Ruane, Garreth and Krist, John and Groff, Tyler},
  booktitle={Techniques and Instrumentation for Detection of Exoplanets IX},
  volume={11117},
  pages={21--36},
  year={2019},
  organization={SPIE}
}

@ARTICLE{Juanola_Parramon_2026,
       author = {{Juanola-Parramon}, Roser and {Chen}, Pin and {Damiano}, Mario and {Gersh-Range}, Jessica and {Groff}, Tyler and {Kasdin}, Jeremy and {Korzh}, Boris and {Krist}, John and {Mamajek}, Eric and {Mazin}, Benjamin and {Morgan}, Rhonda and {Potier}, Axel and {Pueyo}, Laurent and {Quijada}, Manuel and {Sadleir}, John and {Scowen}, Paul and {Stapelfeldt}, Karl and {Stark}, Christopher and {Siegler}, Nick and {Young}, Amber and {Zimmerman}, Neil},
        title = "{Toward a near-ultraviolet coronagraph instrument for the Habitable Worlds Observatory}",
      journal = {Journal of Astronomical Telescopes, Instruments, and Systems},
         year = 2026,
        month = jan,
       volume = {12},
          eid = {015005},
        pages = {015005},
          doi = {10.1117/1.JATIS.12.1.015005},
       adsurl = {https://ui.adsabs.harvard.edu/abs/2026JATIS..12a5005J}
}

@ARTICLE{Konrad_2022,
       author = {{Konrad}, B.~S. and {Alei}, E. and {Quanz}, S.~P. and {Angerhausen}, D. and {Carri{\'o}n-Gonz{\'a}lez}, {\'O}. and {Fortney}, J.~J. and {Grenfell}, J.~L. and {Kitzmann}, D. and {Molli{\`e}re}, P. and {Rugheimer}, S. and {Wunderlich}, F. and {LIFE Collaboration}},
        title = "{Large Interferometer For Exoplanets (LIFE). III. Spectral resolution, wavelength range, and sensitivity requirements based on atmospheric retrieval analyses of an exo-Earth}",
      journal = {Astronomy \& Astrophysics},
         year = 2022,
        month = aug,
       volume = {664},
          eid = {A23},
        pages = {A23},
          doi = {10.1051/0004-6361/202141964},
archivePrefix = {arXiv},
       eprint = {2112.02054},
 primaryClass = {astro-ph.EP},
       adsurl = {https://ui.adsabs.harvard.edu/abs/2022A&A...664A..23K}
}

@ARTICLE{Young_2025,
       author = {{Young}, Amber V. and {Robinson}, Tyler D. and {Krissansen-Totton}, Joshua and {Schwieterman}, Edward W. and {Arney}, Giada and {Lindberg}, Gerrick E. and {Thomas}, Cristina},
        title = "{Modern Earth-like Chemical Disequilibrium Biosignatures are Challenging to Constrain through Spectroscopic Retrievals}",
      journal = {The Astrophysical Journal},
         year = 2025,
        month = jun,
       volume = {986},
       number = {2},
          eid = {206},
        pages = {206},
          doi = {10.3847/1538-4357/addbd7},
archivePrefix = {arXiv},
       eprint = {2505.16231},
 primaryClass = {astro-ph.EP},
       adsurl = {https://ui.adsabs.harvard.edu/abs/2025ApJ...986..206Y}
}

@ARTICLE{Young_2024,
       author = {{Young}, Amber V. and {Robinson}, Tyler D. and {Krissansen-Totton}, Joshua and {Schwieterman}, Edward W. and {Wogan}, Nicholas F. and {Way}, Michael J. and {Sohl}, Linda E. and {Arney}, Giada N. and {Reinhard}, Christopher T. and {Line}, Michael R. and {Catling}, David C. and {Windsor}, James D.},
        title = "{Inferring chemical disequilibrium biosignatures for Proterozoic Earth-like exoplanets}",
      journal = {Nature Astronomy},
         year = 2024,
        month = jan,
       volume = {8},
       number = {1},
        pages = {101-110},
          doi = {10.1038/s41550-023-02145-z},
archivePrefix = {arXiv},
       eprint = {2311.06083},
 primaryClass = {astro-ph.EP},
       adsurl = {https://ui.adsabs.harvard.edu/abs/2024NatAs...8..101Y}
}

@ARTICLE{Young_Decision_Tree_2024,
       author = {{Young}, Amber V. and {Crouse}, Jaime and {Arney}, Giada and {Domagal-Goldman}, Shawn and {Robinson}, Tyler D. and {Bastelberger}, Sandra T.},
        title = "{Retrievals Applied to a Decision Tree Framework Can Characterize Earthlike Exoplanet Analogs}",
      journal = {The Planetary Science Journal},
         year = 2024,
        month = jan,
       volume = {5},
       number = {1},
          eid = {7},
        pages = {7},
          doi = {10.3847/PSJ/ad09b1},
archivePrefix = {arXiv},
       eprint = {2311.07530},
 primaryClass = {astro-ph.EP},
       adsurl = {https://ui.adsabs.harvard.edu/abs/2024PSJ.....5....7Y}
}

@ARTICLE{Titov_2018,
       author = {{Titov}, Dmitrij V. and {Ignatiev}, Nikolay I. and {McGouldrick}, Kevin and {Wilquet}, Val{\'e}rie and {Wilson}, Colin F.},
        title = "{Clouds and Hazes of Venus}",
      journal = {Space Science Reviews},
         year = 2018,
        month = dec,
       volume = {214},
       number = {8},
          eid = {126},
        pages = {126},
          doi = {10.1007/s11214-018-0552-z},
       adsurl = {https://ui.adsabs.harvard.edu/abs/2018SSRv..214..126T}
}

@article{marcq_2018,
  title={Composition and chemistry of the neutral atmosphere of Venus},
  author={Marcq, Emmanuel and Mills, Franklin P and Parkinson, Christopher D and Vandaele, Ann Carine},
  journal={Space Science Reviews},
  volume={214},
  number={1},
  pages={10},
  year={2018},
  publisher={Springer}
}

@article{Gordon_2022_HITRAN,
  title={The HITRAN2020 molecular spectroscopic database},
  author={Gordon, Iouli E and Rothman, Laurence S and Hargreaves, ea RJ and Hashemi, R and Karlovets, Ekaterina Vladimirovna and Skinner, FM and Conway, Eamon K and Hill, Christian and Kochanov, Roman V and Tan, Y and others},
  journal={Journal of quantitative spectroscopy and radiative transfer},
  volume={277},
  pages={107949},
  year={2022},
  publisher={Elsevier}
}

@article{Harman_2015,
  title={Abiotic O2 levels on planets around F, G, K, and M stars: possible false positives for life?},
  author={Harman, CE and Schwieterman, EW and Schottelkotte, James C and Kasting, JF},
  journal={The Astrophysical Journal},
  volume={812},
  number={2},
  pages={137},
  year={2015},
  publisher={The American Astronomical Society}
}

@article{Meadows_2018,
  title={Exoplanet biosignatures: understanding oxygen as a biosignature in the context of its environment},
  author={Meadows, Victoria S and Reinhard, Christopher T and Arney, Giada N and Parenteau, Mary N and Schwieterman, Edward W and Domagal-Goldman, Shawn D and Lincowski, Andrew P and Stapelfeldt, Karl R and Rauer, Heike and DasSarma, Shiladitya and others},
  journal={Astrobiology},
  volume={18},
  number={6},
  pages={630--662},
  year={2018},
  publisher={SAGE Publications Sage CA: Los Angeles, CA}
}

@article{Wordsworth_2014,
  title={Abiotic oxygen-dominated atmospheres on terrestrial habitable zone planets},
  author={Wordsworth, Robin and Pierrehumbert, Raymond},
  journal={The Astrophysical Journal Letters},
  volume={785},
  number={2},
  pages={L20},
  year={2014},
  publisher={The American Astronomical Society}
}

@article{Luger_2015,
  title={Extreme water loss and abiotic O2 buildup on planets throughout the habitable zones of M dwarfs},
  author={Luger, Rodrigo and Barnes, Rory},
  journal={Astrobiology},
  volume={15},
  number={2},
  pages={119--143},
  year={2015},
  publisher={SAGE Publications Sage CA: Los Angeles, CA}
}

@article{Ostberg_2023,
  title={The prospect of detecting volcanic signatures on an exoearth using direct imaging},
  author={Ostberg, Colby M and Guzewich, Scott D and Kane, Stephen R and Kohler, Erika and Oman, Luke D and Fauchez, Thomas J and Kopparapu, Ravi K and Richardson, Jacob and Whelley, Patrick},
  journal={The Astronomical Journal},
  volume={166},
  number={5},
  pages={199},
  year={2023},
  publisher={The American Astronomical Society}
}

@article{Thorngren_2026,
  title={Bayesian model comparison and significance: widespread errors and how to correct them},
  author={Thorngren, Daniel P and Sing, David K and Mukherjee, Sagnick},
  journal={The Astrophysical Journal Supplement Series},
  volume={283},
  number={1},
  pages={10},
  year={2026},
  publisher={The American Astronomical Society}
}

@article{Segura_2003,
  title={Ozone concentrations and ultraviolet fluxes on Earth-like planets around other stars},
  author={Segura, Ant{\'\i}gona and Krelove, Kara and Kasting, James F and Sommerlatt, Darrell and Meadows, Victoria and Crisp, David and Cohen, Martin and Mlawer, Eli},
  journal={Astrobiology},
  volume={3},
  number={4},
  pages={689--708},
  year={2003},
  publisher={SAGE Publications Sage CA: Los Angeles, CA}
}

@article{Rugheimer_2013,
  title={Spectral fingerprints of Earth-like planets around FGK stars},
  author={Rugheimer, Sarah and Kaltenegger, Lisa and Zsom, Andras and Segura, Ant{\'\i}gona and Sasselov, Dimitar},
  journal={Astrobiology},
  volume={13},
  number={3},
  pages={251--269},
  year={2013},
  publisher={SAGE Publications Sage CA: Los Angeles, CA}
}

@article{Kozakis_2022,
  title={Is ozone a reliable proxy for molecular oxygen? I. The O2--O3 relationship for Earth-like atmospheres},
  author={Kozakis, Thea and Mendon{\c{c}}a, Jo{\~a}o M and Buchhave, Lars A},
  journal={Astronomy \& Astrophysics},
  volume={665},
  pages={A156},
  year={2022},
  publisher={EDP Sciences}
}

@article{Hitchcock_1967,
  title={Life detection by atmospheric analysis},
  author={Hitchcock, Dian R and Lovelock, James E},
  journal={Icarus},
  volume={7},
  number={1-3},
  pages={149--159},
  year={1967},
  publisher={Elsevier}
}

@article{lovelock_physical_1965,
	title = {A {Physical} {Basis} for {Life} {Detection} {Experiments}},
	volume = {207},
	copyright = {1965 Nature Publishing Group},
	issn = {1476-4687},
	url = {https://www.nature.com/articles/207568a0},
	doi = {10.1038/207568a0},
	language = {en},
	number = {4997},
	urldate = {2018-08-17},
	journal = {Nature},
	author = {Lovelock, J. E.},
	month = aug,
	year = {1965},
	pages = {568--570},
}

@article{Lovelock1975,
author = {Lovelock, J.E.},
doi = {10.1098/rspb.1975.0051},
issn = {0080-4649},
journal = {Proceedings of the Royal Society of London. Series B. Biological Sciences},
month = {may},
number = {1095},
pages = {167--181},
title = {{Thermodynamics and the recognition of alien biospheres}},
url = {https://royalsocietypublishing.org/doi/10.1098/rspb.1975.0051},
volume = {189},
year = {1975}
}

@ARTICLE{Lippincott_1967,
       author = {{Lippincott}, Ellis R. and {Eck}, Richard V. and {Dayhoff}, Margaret O. and {Sagan}, Carl},
        title = "{Thermodynamic Equilibria in Planetary Atmospheres}",
      journal = {Astrophysical Journal},
         year = 1967,
        month = feb,
       volume = {147},
        pages = {753},
          doi = {10.1086/149051},
       adsurl = {https://ui.adsabs.harvard.edu/abs/1967ApJ...147..753L}
}

@ARTICLE{Nemati_2020,
       author = {{Nemati}, Bijan and {Stahl}, H. Philip and {Stahl}, Mark T. and {Ruane}, Garreth J. and {Sheldon}, Leah J.},
        title = "{Method for deriving optical telescope performance specifications for Earth-detecting coronagraphs}",
      journal = {Journal of Astronomical Telescopes, Instruments, and Systems},
         year = 2020,
        month = jul,
       volume = {6},
          eid = {039002},
        pages = {039002},
          doi = {10.1117/1.JATIS.6.3.039002},
       adsurl = {https://ui.adsabs.harvard.edu/abs/2020JATIS...6c9002N}
}

@article{kass1995bayes,
  title={Bayes factors},
  author={Kass, Robert E and Raftery, Adrian E},
  journal={Journal of the american statistical association},
  volume={90},
  number={430},
  pages={773--795},
  year={1995},
  publisher={Taylor \& Francis}
}

@book{jeffreys1998theory,
  title={The theory of probability},
  author={Jeffreys, Harold},
  year={1998},
  publisher={OuP Oxford}
}

@article{schwarz_1978,
  title={Estimating the dimension of a model},
  author={Schwarz, Gideon},
  journal={The annals of statistics},
  pages={461--464},
  year={1978},
  publisher={JSTOR}
}
\bibliographystyle{spiejour}   





\end{spacing}
\end{document}